\documentclass[aps,pra,amsmath,amssymb,floatfix,twocolumn,amsmath,superscriptaddress,twocolumn,nofootinbib,tighten,letterpaper]{revtex4-1}
\usepackage{lmodern}
\usepackage[T1]{fontenc}
\usepackage[latin9]{inputenc}
\usepackage{float}
\usepackage[normalem]{ulem}
\usepackage{amsmath}
\usepackage{amssymb}
\usepackage{graphicx}
\usepackage{wasysym}
\usepackage{color}
\usepackage{bm}
\usepackage{hyperref}
\hypersetup{pdfpagemode=UseNone}

\DeclareMathOperator{\Tr}{Tr}
\DeclareMathOperator{\sgn}{sgn}

\DeclareMathOperator{\diag}{diag}

\newcommand{\bsub}{\begin{subequations}}
\newcommand{\esub}{\end{subequations}}

\newcommand{\ord}[1]{\bm{\mathit{O}}\left(#1\right)}

\newcommand{\vex}[1]{\bm{\mathrm{#1}}}

\newcommand{\e}{E}
\newcommand{\hh}{\hat{h}}
\newcommand{\Nabla}{\bm{\nabla}}
\newcommand{\tauh}{\hat{\tau}}

\newcommand{\sigh}{\hat{\sigma}}

\newcommand{\T}{{\mathsf{T}}}
\newcommand{\nuqh}{\nu_{\scriptscriptstyle{\mathsf{QHPT}}}}
\newcommand{\Ms}{\hat{M}_{\mathsf{S}}}
\newcommand{\pup}[1]{{\scriptscriptstyle{({#1})}}}
\newcommand{\Stop}{S_{\mathsf{top}}}

\newcommand{\vnn}{\vex{\delta}}
\newcommand{\vnnn}{\tilde{\vex{\delta}}}

\begin{document}

\title{Spectrum-wide quantum criticality at the surface of class AIII topological phases:\\
An ``energy stack'' of integer quantum Hall plateau transitions}

\author{Bj\"orn Sbierski}
\affiliation{Department of Physics, University of California, Berkeley, CA 94720, USA}
\author{Jonas F. Karcher}
\affiliation{Institut f\"ur Nanotechnologie, Institut f\"ur QuantenMaterialien und Technologien and Institut f\"ur Theorie
der Kondensierten Materie, Karlsruhe Institute of Technology, 76021
Karlsruhe, Germany}
\affiliation{Department of Physics and Astronomy, Rice University, Houston, Texas 77005, USA}
\author{Matthew S. Foster}
\affiliation{Department of Physics and Astronomy, Rice University, Houston, Texas 77005, USA}
\affiliation{Rice Center for Quantum Materials, Rice University, Houston, Texas 77005, USA}

\date{\today}

\begin{abstract}
In the absence of spin-orbit coupling, the conventional dogma of Anderson localization
asserts that all states localize in two dimensions, with a glaring exception: the quantum Hall
plateau transition (QHPT). In that case, the localization length diverges
and interference-induced quantum-critical spatial fluctuations appear at all length scales.
Normally QHPT states occur only at isolated energies; accessing them therefore requires
fine-tuning of the electron density or magnetic field. 
In this paper we show that QHPT states can be realized throughout an energy continuum, 
i.e.\ as an ``energy stack'' of critical states wherein each state in the stack exhibits QHPT phenomenology.
The stacking occurs \emph{without fine-tuning} at the surface of a class AIII topological phase,
where it is protected by U(1) and (anomalous) chiral or time-reversal symmetries. 
Spectrum-wide criticality is diagnosed by comparing numerics to universal results for the longitudinal Landauer 
conductance and wave function multifractality at the QHPT. Results are obtained from an effective 2D surface field theory 
and from a bulk 3D lattice model. We demonstrate that the stacking of quantum-critical QHPT states
is a robust phenomenon that occurs for AIII topological phases with both odd and even winding numbers. 
The latter conclusion may have important implications for the still poorly-understood logarithmic conformal field theory believed to describe the QHPT. 
\end{abstract}

\maketitle

\section{Introduction}

\begin{figure}[b!]
\includegraphics{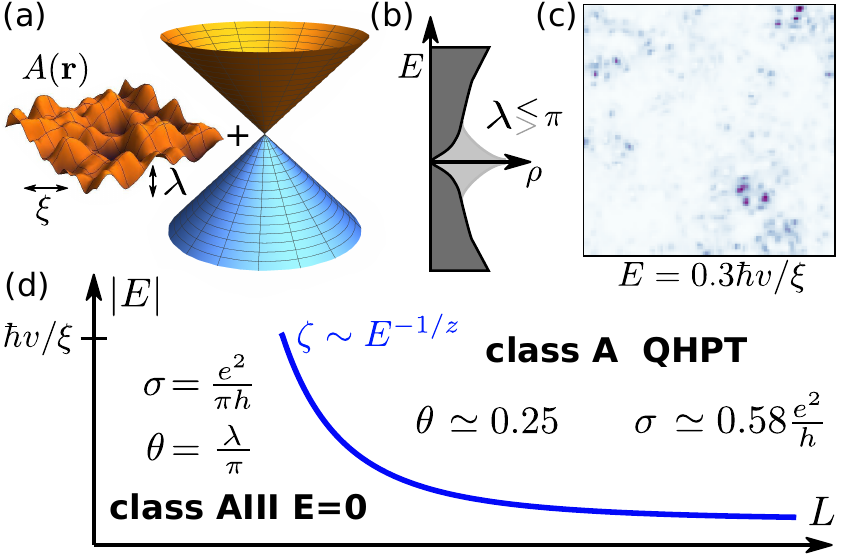}
\caption{\label{fig:AIII_summary}
Summary of main results. 
(a) We consider a 2D Dirac node in the presence of quenched vector potential disorder $A(\mathbf{r})$ of correlation length 
$\xi$ and strength $\lambda$, modeling the surface state of a topological phase in symmetry class AIII. 
It is well known that the zero energy field theory is critical while finite energies $E$ feature a power-law density of states 
(b) and an emergent length scale $\zeta \sim E^{-1/z}$. 
(c) By studying the real-space characteristics of finite energy eigenstates (conductivity $\sigma$ and the curvature of the 
multifractality spectrum $\theta$), we identify $\zeta$ as a crossover scale beyond which criticality of the quantum Hall 
plateau transition (QHPT) type emerges.
Our results imply that the finite-energy states form a ``stack'' of critical QHPT wave functions;
this is in sharp contrast to the conventional expectation of Anderson localization \cite{Ludwig1994}.
We also show that this unusual energy-continuum of critical states is stable under the addition of a second Dirac node.}
\label{Fig--Summary}
\end{figure}

Non-interacting topological quantum phases of matter feature robust gapless edge or surface states
\cite{KaneHasan,TSCRev1,BernevigBook}. Remarkably, these states are protected from Anderson localization
\cite{SRFL2008,Essin2015,Essin2011}, defying the central dogma of conventional localization physics according 
to which all states localize in one dimension, as do most states in two dimensions \cite{Evers2008}. 
An exception occurs for 2D systems with strong spin-orbit coupling 
and sufficiently weak disorder; in this case Anderson localization occurs near the band edges, but states
near the band center can remain delocalized due to weak antilocalization \cite{LeeRamakrishnan1985}. 
By contrast, for the chiral [helical] edge states of the quantum Hall effect [2D topological insulators (TIs)],
as well as the isolated 2D Dirac cone at the boundary of a 3D TI, \emph{all states} spanning the bulk 
gap are prevented from localizing. 

While the protection of chiral and helical 1D edge states is easily understood as the absence of elastic 
backscattering \cite{KaneMele2005,Xie2016,KaneHasan,TSCRev1},
the effects of disorder on 2D surface states of bulk topological phases is more subtle. 
Indeed, although pure backscattering is suppressed for the single Dirac fermion cone that forms at the boundary 
of the simplest 3D TI, elastic impurity scattering does occur at all other angles. 
The protection
of the 2D surface states throughout the bulk energy gap is understood from the combination of 
weak antilocalization, as well as a $\mathbb{Z}_2$ topological term in the effective field theory 
for the surface with quenched disorder (a nonlinear sigma model in the symplectic class)
\cite{Bardarson2007,Nomura2007,Ryu2007,Ostrovsky2007,Evers2008,Konig2014}. 

In this paper, we consider the effects of disorder on 2D Dirac surface states of class AIII topological phases, 
at energies throughout the bulk gap. (See Table~\ref{table:10-fold-way} for a review of the 10-fold classification).
 In the absence of strong interactions \cite{Wang2014}, class AIII is characterized 
by a $\mathbb{Z}$-valued winding number $\nu$ in three dimensions, reflecting the number of 2D Dirac nodes on the surface. 
In this case, due to the absence of full spin-orbit coupling, protection from Anderson localization cannot be attributed to weak antilocalization. 
Class AIII could be realized either as a topological insulator on a bipartite lattice,
with complex hopping that preserves sublattice symmetry but breaks time-reversal, 
or as a time-reversal invariant, spin-triplet topological superconductor (TSC) that preserves
a U(1) remnant of spin SU(2) symmetry \cite{Foster2008,SRFL2008,Hosur2010,TSCRev2}.

The effects of quenched disorder for class AIII surface states
at energies close to zero are well known; 
here energy is measured relative to the surface Dirac point 
and the disorder comes in the form of a random vector potential.
Topologically-protected continuum Dirac models 
were originally studied in the contexts of  
the quantum Hall effect \cite{Ludwig1994} and $d$-wave superconductor quasiparticles
\cite{Nersesyan1994,Mudry1996,Caux1996,Bhaseen2001,AltlandSimonsZirnbauer2002,DellAnna2007}.
The zero-energy states always escape Anderson localization \cite{Ludwig1994,Nersesyan1994}. 
Instead, these states are quantum critical, exhibiting strongly inhomogeneous
spatial fluctuations on all length scales. These fluctuations can be
characterized by the multifractal spectrum of wave function intensities \cite{Evers2008}, 
which is known exactly in these models \cite{Ludwig1994,Mudry1996,Caux1996,Chamon1996}. 
Two other key attributes include a critical low-energy density of states $\rho(\e)$
that vanishes or diverges as $\e \rightarrow 0$,
and a precisely quantized longitudinal conductivity at zero energy, 
independent of disorder \cite{Ludwig1994,Tsvelik1995,Ostrovsky2006,Evers2008}
and (at zero temperature) interactions \cite{Xie2015}.

\begin{table*}[!t]
{\renewcommand{\arraystretch}{1.2}
\begin{tabular*}{\textwidth}{l @{\extracolsep{\fill}} ccccccll}
\hline
\hline
	Class
& 
	$T$
&
	$P$
& 
	$S$
&
	Spin sym. 
&
	$d = 2$
&
	$d = 3$
& 
	Topological realization	
& 
	Replicated fermion NLsM	
\\
\hline
C 				& 0 & -1 & 0 	& SU(2) & $2 \mathbb{Z}$ 	& - 			& SQHE (2D $d+id$ TSC)		&	 ${\mathrm{Sp}(4n)}\,/\,{\mathrm{U}(2n)}$
\\
A {\tiny{(unitary)}}		& 0 & 0 & 0 	& U(1) 	& $\mathbb{Z}$ 		& - 			& IQHE 				&	${\mathrm{U}(2n)}\,/\,{\mathrm{U}(n) \otimes \mathrm{U}(n)}$
\\
D 				& 0 & +1 & 0 	& - 	& $\mathbb{Z}$ 		& - 			& TQHE (2D $p+ip$ TSC) 		&	${\mathrm{O}(2n)}\,/\,{\mathrm{U}(n)}$ 
\\
\phantom{0}
\\		
CI 				& +1 & -1 & 1 	& SU(2) & -		 	& $2 \mathbb{Z}$ 	& 3D TSC 			& 	${\mathrm{Sp}(4n) \otimes \mathrm{Sp}(4n)}\,/\,{\mathrm{Sp}(4n)}$
\\
AIII 				& 0 & 0 & 1	& U(1) 	& - 			& $\mathbb{Z}$ 	 	& 3D TSC, chiral TI		& 	${\mathrm{U}(2n) \otimes \mathrm{U}(2n)}\,/\,{\mathrm{U}(2n)}$
\\
DIII 				& -1 & +1 & 1 	& - 	& $\mathbb{Z}_2$	& $\mathbb{Z}$ 	& 3D TSC ($^3$He-$B$)			& 	${\mathrm{O}(2n) \otimes \mathrm{O}(2n)}\,/\,{\mathrm{O}(2n)}$ 
\\
\phantom{0}
\\
AI {\tiny{(orthogonal)}}	& +1 & 0 & 0 	& SU(2) & -		 	& - 			& -		 		& 	${\mathrm{Sp}(4n)}\,/\,{\mathrm{Sp}(2n) \otimes \mathrm{Sp}(2n)}$
\\
AII {\tiny{(symplectic)}} 	& -1 & 0 & 0	& - 	& $\mathbb{Z}_2$ 	& $\mathbb{Z}_2$ 	& 2D, 3D TIs 			& 	${\mathrm{O}(2n) }\,/\,{\mathrm{O}(n) \otimes \mathrm{O}(n)}$
\\
\phantom{0}
\\
BDI 				& +1 & +1 & 1 	& SU(2)	& -		 	& - 			& -		 		& 	${\mathrm{U}(2n)}\,/\,{\mathrm{Sp}(2n)}$
\\
CII 			 	& -1 & -1 & 1	& - 	& -		 	& $\mathbb{Z}_2$ 	& 3D chiral TI 			& 	${\mathrm{U}(2n) }\,/\,{\mathrm{O}(2n)}$
\\
\hline 
\hline
\end{tabular*}
}
\caption{The 10-fold way classification for strong (fully gapped), 
$d$-dimensional symmetry-protected topological phases of 
fermions, i.e.\ 
topological insulators (TIs) and topological superconductors (TSCs) \cite{SRFL2008}. 
The 10 classes are defined by different combinations of the three \emph{effective} discrete symmetries 
$T$ (time-reversal), 
$P$ (particle-hole), 
and 
$S$ (chiral or sublattice).
For a $d$-dimensional bulk, any deformation of the clean band structure that preserves $T$, $P$, and $S$ 
and does not close a gap preserves the topological winding number.
For a $(d-1)$-dimensional edge or surface theory, the equivalent statement is that any static
deformation of the surface (quenched disorder) that preserves $T$, $P$, and $S$ also preserves the 
``topological protection'' against Anderson localization. 
Of particular interest here are classes C, A, D on one hand, and classes CI, AIII, DIII on the other. 
Classes C, A, and D are topological in $d = 2$, and describe the spin (SQHE), integer (IQHE), and thermal (TQHE) quantum Hall effects;
all three can be realized as TSCs with broken $T$. 
Classes CI, AIII, and DIII are topological in $d = 3$, and can describe 3D time-reversal-invariant TSCs. 
(In this case, the physical time-reversal symmetry appears as the \emph{effective} chiral symmetry $S$ in the table \cite{SRFL2008,Foster2008}.)
The column ``spin sym.'' denotes the amount of spin SU(2) symmetry preserved for TSC realizations of these 6 classes. 
In this work, we show that the finite-energy surface states of the 3D class AIII topological phase 
appear to form a ``stack'' of quantum-critical, Anderson delocalized wave functions. 
We provide evidence that each state in the stack is statistically identical on large length scales, and
corresponds precisely to the \emph{topological quantum phase transition} in class A (the integer quantum Hall plateau transition).
Previous numerical results also uncovered a gap-spanning ``stack'' of critical states at the surface of the class CI 
TSC, which match the statistics of the class C spin quantum Hall plateau transition \cite{Ghorashi2018}.
A third study \cite{Ghorashi2019} revealed spectrum-wide criticality at the surface of class DIII,
conjectured to be related to the thermal quantum Hall plateau transition in class D.
Thus the numerical results presented in this paper and in Refs.~\cite{Ghorashi2018,Ghorashi2019} 
appear to connect the quantum Hall plateau transitions in classes C, A, and D to gap-spanning
stacks of critical states at the surfaces of class CI, AIII, and DIII 3D topological phases. 
The conclusion of gap-spanning surface criticality, locked to plateau transitions in classes C, A, D, 
is in sharp contrast to the conventional expectation for finite-energy 2D surface states
in classes CI, AIII, DIII. 
The conventional expectation is that finite energy always breaks the defining chiral $S$ symmetry,
producing a standard Wigner-Dyson class,
so that \cite{Evers2008}
CI 	$\rightarrow$ AI 	(Anderson localized),
AIII 	$\rightarrow$ A 	(Anderson localized),
and
DIII 	$\rightarrow$ AII	(Anderson localized or weak antilocalization) 
(see however \cite{SigmaModelManifoldNote}).
}
\label{table:10-fold-way}
\end{table*}

Despite this plethora of exact results at zero energy, very little 
was known about the character of finite-energy states or of the finite-temperature response.  
Ludwig \emph{et al}.\ \cite{Ludwig1994}
argued that all finite-energy states of a 2D Dirac model with one node
(corresponding to the bulk winding number $\nu = 1$) 
should Anderson localize.
Later, Ostrovsky \emph{et al}.\ \cite{Ostrovsky2007} instead conjectured 
that finite-energy states in this model could escape
localization via a remarkable route: each of these states would
sit exactly at the integer quantum Hall plateau transition (QHPT). 
The QHPT is a quantum phase transition that divides Anderson topological-insulating quantum Hall plateaux;
it is known to also be quantum critical, with apparently universal
multifractal spectra and average longitudinal conductivity \cite{Huckestein1995,Huo1993,Cho1997,Wang1998,Schweitzer2005,Jovanovic1998,Evers2008}.   
Ostrovsky \emph{et al}.\ suggested however that this would only hold very close
to the Dirac point \cite{Ostrovsky2007}. Moreover, the derivation would seem to imply an 
``even-odd'' effect, similar to the Haldane conjecture for half-integer versus integer Heisenberg spin chains. 
Specifically, finite-energy class AIII topological surface states with odd winding numbers are predicted
to exhibit QHPT criticality, while those with even winding numbers would simply localize 
\cite{Ostrovsky2007,Konig2014,Essin2015}.

In this paper, we present strong numerical evidence that finite-energy surface states throughout the bulk 
gap of a class AIII topological phase exhibit QHPT criticality 
beyond the crossover length scale $\zeta \sim \e^{-1/z}$ of the zero energy critical theory, see Fig.~\ref{Fig--Summary} for a graphical summary.   
Moreover, we find no evidence
of an even-odd effect. Calculations are performed for winding numbers $\nu = \{1,2\}$ using two different
model types:
(1) effective low-energy 2D surface field theories
and
(2) surface states of 3D topological lattice models in the slab geometry. 
For the continuum theories, we compute the energy-resolved 
Landauer conductivity,
Landauer conductance distribution,
multifractal spectrum,
and 
density of states. 
We observe the crossover of the conductivity between the exact zero-energy class AIII result
and the known average conductivity of the QHPT \cite{Huo1993,Cho1997,Wang1998,Schweitzer2005}; 
concomitantly we observe QHPT multifractality.
We demonstrate the crossover between two copies of the $\nu = 1$ model 
to $\nu = 2$ as a function of internode scattering. 
For square systems, we obtain a Landauer conductance distribution that scales towards the known 
QHPT result \cite{Jovanovic1998} with increasing system size for both $\nu = 1,2$.
Our results confirm and significantly extend a previous glimpse of QHPT multifractality at
finite energy reported in Ref.~\cite{Chou2014}. 
For the lattice models, we provide evidence that states throughout the gap avoid localization, 
and that finite-energy states exhibit multifractal spectra consistent with the QHPT. 
We summarize our main findings in Fig.~\ref{Fig--Summary}. 

We conclude that class AIII appears to solve the problem of topologically protecting its surface state spectrum
throughout the bulk gap by forming an \emph{energy continuum} of QHPT states. 
This is surprising, 
given that in the 2D quantum Hall effect, critical states appear only at isolated
energies; accessing them in experiment requires fine-tuning of the electron density or magnetic field to a quantum critical point. 
The phenomenon of spectrum-wide criticality observed here is very unusual, but has been previously 
reported in two other cases: models for surface states of class CI and DIII topological
superconductors (TSCs) \cite{Ghorashi2018,Ghorashi2019}. The finite-energy
states in class CI \cite{Ghorashi2018}
were found to exhibit critical statistics consistent with the class C \emph{spin} 
quantum Hall plateau transition \cite{SQHPT-1_Kagalovsky1999,SQHPT-2_Gruzberg1999,SQHPT-3_Senthil1999,SQHPT-4_Cardy2000,SQHPT-5_Beamond2002,SQHPT-6_Evers2003,SQHPT-7_Mirlin2003,Evers2008}.
A single surface Majorana fermion cone is predicted to occur at the boundary 
of a class DIII TSC, such as the candidate material Cu$_x$Bi$_2$Se$_3$ \cite{TSCRev1,TSCRev3}.
In the presence of disorder, finite-energy class DIII Majorana surface states 
also appear to exhibit robust, universal criticality \cite{Ghorashi2019}, conjectured to be related
to the thermal quantum Hall plateau transition in class D \cite{D-1_SenthilFisher2000,D-2_BocquetZirnbauer2000,D-3_ReadLudwig2000,D-4_Gruzberg2001,D-5_Chalker2001,D-6_Mildenberger2007,D-7_Laumann2012}.
The 10-fold way symmetry classification is reviewed in Table~\ref{table:10-fold-way},
with an emphasis on the implied connections between surface states of the 3D topological classes CI, AIII, DIII, 
and 
2D topological phase transitions in classes C, A, D.

Together, the results obtained in this paper and in Refs.~\cite{Ghorashi2018,Ghorashi2019} imply that the ordinary 3D class AII, 
$\mathbb{Z}_2$ topological insulator is actually the exceptional case. 
The metallic surface states of a class AII TI do not exhibit universal criticality; instead, the classical surface conductivity 
at each energy in the gap is determined by the impurity density and the density of states, and this nonuniversal, energy-dependent
value is continuously enhanced at larger distances due to weak antilocalization \cite{Bardarson2007,Nomura2007,Ryu2007,Ostrovsky2007,Evers2008}.  
On the other hand, 
three of the five \cite{SRFL2008,TSCRev2} topological classes in 3D
exhibit spectrum-wide quantum criticality. 
There appears to be an unexpected, integral connection between $\mathbb{Z}$-graded topological phases 
in two and three spatial dimensions: the quantum phase transitions in the former (classes C, A, D) are bundled into gap-spanning stacks
of surface eigenstates in the latter (classes CI, AIII, DIII). Our results should have a deeper topological underpinning,
which however remains yet to be uncovered. 

Our results may also have implications for the theory of the class A integer QHPT itself.
The transition has been extensively studied numerically \cite{Huckestein1995,Evers2008}, with universal signatures consistent 
with a conformally invariant critical point. Despite 30 years of effort, very little is known
about this critical point analytically, which is believed to be governed by a logarithmic conformal field theory (LCFT) \cite{Cardy2013}.
A recent proposal by Zirnbauer \cite{Zirnbauer2019} in fact argues that the QHPT is
itself governed by a zero-energy, class AIII theory with winding number $\nu = 4$. 
An additional special feature is that the purely marginal abelian disorder strength 
(a generic feature for class AIII theories, 
reviewed below) is fine-tuned to a special value so as to render the density of states non-critical. 
A very surprising consequence of this proposal is that the correlation length exponent $\nuqh$ is actually
predicted to be \emph{infinite}; the claim being that existing numerical studies showing $\nuqh$
in the range between $2.3$--$2.6$
\cite{Huckestein1995,Evers2008,Slevin2009,Obuse2010,Amado2011,Dahlhaus2011,Fulga2011,Obuse2012,Slevin2012,Nuding2015,Gruzberg2017}
are beset by finite-size effects. 
We note that for the finite-energy class AIII surface states that exhibit QHPT critical statistics,
the exponent $\nuqh$ does not play a role because all states are delocalized.

If Zirnbauer's proposal is correct, then our numerical results suggest that \emph{both} the zero- and 
finite-energy states of the class AIII bulk topological phase with the special winding number $\nu = 4$ are governed
by the \emph{same} conformal field theory. In this scenario, the energy perturbation serves only
to fine-tune the abelian disorder strength to the correct value so as to achieve QHPT criticality. 
At least this proposal should be falsifiable using the class AIII surface state theory, 
which is a relatively well-understood LCFT
(a Wess-Zumino-Novikov-Witten model \cite{Nersesyan1994,Mudry1996,Caux1996,Bhaseen2001,Foster2014}).

\subsection{Outline}

This paper is organized as follows. 
In Sec.~\ref{sec:model}, we transcribe a disordered Dirac Hamiltonian for 
two nodes. We enumerate topological (anomalous) and regular symmetry operations
and identify $\nu = 1,2$ class AIII surface state models. 
We also define two non-topological models that will be studied for comparison:
a Dirac version of the Gade ``sublattice hopping'' class AIII model \cite{GadeWegner1991,Gade1993,Guruswamy2000},
and a topologically trivial version of spinless graphene in the unitary class A. 
We summarize key known results for these models, as well as for the quantum Hall plateau transition. 

In Sec.~\ref{sec:nu=1}, we present numerical results for the $\nu = 1$ topological
AIII model at zero and finite energies. These include the density of states,
the multifractal spectrum for wave function and local density of states fluctuations,  
the Landauer conductivity,
and the distribution function for the Landauer conductance.
In Sec.~\ref{sec:nu=2}, we repeat this analysis for the $\nu =2$ topological surface
state model. 
In Sec.~\ref{sec:nontop}, we present analogous results for the non-topological 
Dirac models defined in Sec.~\ref{sec:model}.

Results for multifractal spectra of surface states obtained from a 3D topological
lattice model in the slab geometry, with bulk winding numbers $\nu = \{1,2\}$, are presented in Sec.~\ref{sec:lattice}.
We discuss our results in the context of similar findings in classes CI and DIII \cite{Ghorashi2018,Ghorashi2019}
in Sec.~\ref{sec:disc}, and in particular highlight possible implications for the analytical
understanding of the logarithmic conformal field theory \cite{Cardy2013}
governing the QHPT. The latter
is explored in light of a very recent proposal by Zirnbauer \cite{Zirnbauer2019} linking 
the QHPT to an effective AIII surface state model with winding number $\nu = 4$.


\section{2D Dirac models, symmetry classes, and topological surface theories \label{sec:model}}

The low-energy physics of the class AIII topological surface states that we study in this
paper in Secs.~\ref{sec:nu=1} and \ref{sec:nu=2}, along with two non-topological models employed for comparison
(Sec.~\ref{sec:nontop}), 
can be encoded in a 2D, two-valley Dirac-fermion Hamiltonian. 
The most generic model can be considered the low-energy effective field theory for spinless graphene, 
with the two cones corresponding to the $K$ and $K'$ valleys. 
The Hamiltonian is \cite{Evers2008}
\begin{align}
	H
	=&\,
	\int d^2\vex{r} 
	\,
	\psi^\dagger
	\,
	\hh
	\,
	\psi,
	\quad
	\hh
	\equiv
	\hh_0
	+
	\hh_A
	+
	\hh_B
	+
	\hh_C,
\end{align}
where 
\bsub\label{hDef}
\begin{align}
\label{h0Def}
	\hh_0 
	=&\,
	-i \, v \, \sigh_{\bar{a}} \, \partial_{\bar{a}},
\\
\label{hADef}
	\hh_A
	=&\,	
	m_{\bar{a}} \, \sigh_3 \, \tauh_{\bar{a}} 
	+
	A_{\bar{a},3} \, \sigh_{\bar{a}} \, \tauh_3,
\\
\label{hBDef}
	\hh_B
	=&\,	
	V_{0}
	+
	m_{3} \, \sigh_3 \, \tauh_{3} 
	+
	A_{\bar{a},\bar{b}} \, \sigh_{\bar{a}} \, \tauh_{\bar{b}},
\\
\label{hCDef}
	\hh_C
	=&\,
	V_{\bar{a}} \, \tauh_{\bar{a}} 
	+
	V_{3} \, \tauh_{3}
	+
	m_0 \, \sigh_3
	+
	A_{\bar{a},0} \, \sigh_{\bar{a}}, 
\end{align}
\esub
where $\bar{a} \in \{1,2\}$ and repeated indices are summed. 
The parameter $v$ is the Fermi velocity, while 
the 16 potentials $\{m_{\mu},A_{1,\mu},A_{2,\mu},V_{\mu}\}$ ($\mu \in \{0,1,2,3\}$) 
encode quenched disorder \cite{AleinerEfetov2006}.  
In the context of graphene, the four-component Dirac spinor is defined via  
$
	\psi^\T \equiv \begin{bmatrix} c_{A,K} & c_{B,K} & c_{B,K'} & - c_{A,K'} \end{bmatrix},
$
where $\T$ denotes the matrix transpose, 
and where 
$c_{A,K}$ ($c_{B,K'}$) annihilates an electron on sublattice $A$ ($B$) in valley $K$ ($K'$). 
The Pauli matrices $\{\sigh_{1,2,3}\}$ ($\{\tauh_{1,2,3}\}$) act on sublattice (valley) space, 
so that $\sigh_3 \Rightarrow \diag\{1,-1,1,-1\}$ and $\tauh_3 \Rightarrow \diag\{1,1,-1,-1\}$. 

For spinless fermions hopping on the honeycomb lattice, three key symmetries are 
$T^2 = +1$ time-reversal, $P^2 = +1$ particle-hole, and sublattice $S$ symmetry. The latter
is the product of $T$ and $P$. 
With respect to the single-particle Hamiltonian $\hh$, these symmetries are encoded in the conditions \cite{FosterAleiner2008}
\bsub\label{Latt_Sym}
\begin{align}
T:&\,&
	\sigh_2 \, \tauh_2 \, \hh^* \, \sigh_2 \, \tauh_2 =&\, \hh, &&
\\
P:&\,&
	- \sigh_1 \, \tauh_1 \, \hh^\T \, \sigh_1 \, \tauh_1 =&\, \hh, && 
\\
S:&\,&
	- \sigh_3 \, \tauh_3 \, \hh \, \sigh_3 \, \tauh_3 =&\, \hh, &&
\label{SDef}
\end{align}
\esub 
where $*$ denotes the complex conjugate. 
The kinetic Hamiltonian in Eq.~(\ref{h0Def}) as well as the bilinear perturbations in $\hh_A$ [Eq.~(\ref{hADef})] are invariant under $T$, $P$, and $S$. 
Those in $\hh_B$ [Eq.~(\ref{hBDef})] are invariant under $T$, but not $P$ or $S$,
while those in $\hh_C$ [Eq.~(\ref{hCDef})] are odd under $T$; some are invariant under $P$ or $S$. 

In the graphene context, $\{m_{1,2}\}$, $m_3$, and $m_0$ respectively denote Kekul\'e, CDW, and Haldane 
masses \cite{GrapheneReview2011,BernevigBook}. 
The particle-hole symmetric vector potentials $\{A_{\bar{a},3}\}$ can encode elastic strain (via pseudomagnetic fields) \cite{Peres2010}. 
The scalar potential is $V_0$, while real magnetic fields can be encoded in  $\{A_{\bar{a},0}\}$.

\subsection{Non-topological class A and AIII models \label{sec:nontop_exp}}

In the absence of $T$, $P$, and $S$, the Dirac Hamiltonian in Eq.~(\ref{hDef}) resides in the unitary class A.
If we take all 16 bilinear perturbations to be Gaussian random variables with zero average (including a vanishing 
average Haldane/Chern insulator mass $\overline{m_0} = 0$), then this theory realizes a topologically trivial 
Anderson insulator, with localized states at all energies \cite{Ostrovsky2006,Evers2008}. 

We can get a purely two-dimensional, \emph{nontopological-surface-state} version of class AIII by 
imposing only sublattice symmetry $S$. 
In the honeycomb lattice model, 
sublattice symmetry is implemented via
\begin{align}\label{SDefLatt}
	c_{A}(\vex{r}_A) \rightarrow c_{A}^\dagger(\vex{r}_A), 
	\;\;
	c_{B}(\vex{r}_B) \rightarrow - c_{B}^\dagger(\vex{r}_B),
	\;\;
	i \rightarrow -i.
\end{align}
This is a symmetry for real or complex nearest-neighbor hopping on any bipartite lattice at half-filling \cite{GadeWegner1991,Gade1993}. 
The allowed bilinear perturbations implied by Eq.~(\ref{SDef}) are 
$\{m_1,m_2,A_{1,3},A_{2,3},V_1,V_2,A_{1,0},A_{2,0}\}$ \cite{Guruswamy2000}. 
The vector potentials act independently on the two nodes, 
while the scalar and mass potentials scatter between them. 
This class AIII Hamiltonian can alternatively describe Bogoliubov-de Gennes
quasiparticles in a 2D $p_x$-wave, spin-triplet  (``polar'' \cite{Ho1998,Ohmi1998}) superconductor \cite{Foster2008}. 

At the band center 
($\equiv$ zero energy), 
sublattice symmetry 
introduces additional Goldstone modes in the nonlinear sigma model target manifold relative to ordinary metals. 
As a result, states near zero energy
evade Anderson localization, as originally shown by Gade and Wegner \cite{GadeWegner1991,Gade1993}.
Localization is still possible for strong random hopping \cite{Koenig2012}.  On the other hand,
at energies away from zero we expect eigenstates to reside in an ordinary Wigner-Dyson class. 
For a system possessing only sublattice symmetry, the finite-energy states are expected
to reside in the unitary class A, which localizes without fine-tuning to the quantum Hall plateau transition. 
(See Sec.~\ref{sec:sigma} for a precise nonlinear sigma model formulation in the topological case.) 
Thus all finite-energy states of a non-topological class AIII model in two dimensions 
are expected to localize.

\subsection{Topological AIII models with $\nu \in \{1,2\}$ \label{sec:top_exp}}

Classes CI, AIII, and DIII can describe quasiparticles in time-reversal invariant superconductors,
with SU(2), U(1), or no spin symmetry, respectively \cite{SRFL2008,TSCRev2}.
For any Bogoliubov-de Gennes Hamiltonian, physical time-reversal invariance can always be
represented as an effective ``chiral'' symmetry, i.e.\ the condition that some matrix $\Ms$ 
anticommutes with the single-particle Hamiltonian matrix $\hh$ \cite{Foster2014}.
The sublattice symmetry in Eq.~(\ref{SDef}) is such a chiral condition. 

There are in fact \emph{two} unitarily inequivalent types of chiral symmetry
available for 2D Dirac fermions \cite{BernardLeClair2002}. The sublattice symmetry in Eq.~(\ref{SDef}) 
defines a 2D class AIII Dirac theory with mass, scalar potential, and vector potential perturbations. 
Since one can gap out the system with a nonzero average mass without breaking the defining chiral symmetry, 
this cannot be a topological surface state theory. On the other hand, for the same 2-node Dirac
Hamiltonian in Eq.~(\ref{hDef}), we can introduce an ``anomalous'' version of the chiral symmetry,
\begin{align}\label{STopDef}
	\Stop:&\,&
	- \sigh_3 \, \hh \, \sigh_3 =&\, \hh. &&
\end{align}
This differs from Eq.~(\ref{SDef}) due to the absence of a grading in valley space. 
In the context of a class AIII topological superconductor (TSC), Eq.~(\ref{STopDef}) represents
physical time-reversal symmetry realized at the boundary of the sample \cite{SRFL2008,Foster2014,TSCRev2}. 
In a chiral bulk topological insulator, it is the surface projection of the \emph{bulk} sublattice symmetry \cite{Hosur2010}.
Eq.~(\ref{STopDef}) is anomalous because it cannot be realized via a local symmetry transformation 
in a strictly 2D lattice model [unlike Eq.~(\ref{SDef}), which is the continuum version of the lattice
operation in Eq.~(\ref{SDefLatt})]. 

Imposing Eq.~(\ref{STopDef}) on the Dirac model, only vector potential perturbations are allowed. 
I.e., the Hamiltonian in Eq.~(\ref{hDef}) reduces to 
\begin{align}\label{hAIII2}
	\hh_{\mathsf{AIII}}^{\pup{2}}
	\equiv
	\sigh_{\bar{a}}
	\left[
	-i \, v \, \partial_{\bar{a}} 
	+
	A_{\bar{a},0}(\vex{r})
	+
	A_{\bar{a},i}(\vex{r})
	\,
	\tauh_i
	\right],
\end{align}
where the repeated indices $\bar{a} \in \{1,2\}$ and $i \in \{1,2,3\}$ are summed. 
Eq.~(\ref{hAIII2}) describes 2D Dirac fermions perturbed by quenched
random U(1) ($A_{\bar{a},0}$) and SU(2) ($A_{\bar{a},i}$) vector potentials.
If we further suppress scattering between the two nodes, we get
two copies of the simpler single-node theory
\begin{align}\label{hAIII1}
	\hh_{\mathsf{AIII}}^{\pup{1}}
	\equiv
	\sigh_{\bar{a}}
	\left[
	-i \, v \, \partial_{\bar{a}} 
	+
	A_{\bar{a},0}(\vex{r})
	\right].
\end{align}
In this case only abelian vector potential disorder appears. 
Eqs.~(\ref{hAIII2}) and (\ref{hAIII1}) can be realized
as surface state theories for class AIII TSCs with
bulk winding numbers $\nu = 2$ and $\nu = 1$, respectively. 
Alternatively, two independent copies of Eq.~(\ref{hAIII1}) 
obtain from the 2D class AIII Gade-Dirac model defined by Eq.~(\ref{SDef}),
when internode scattering (mediated by $\{m_1,m_2,V_1,V_2\}$)
is suppressed by hand. Similar fine-tuning of impurity amplitudes
in a model for 2D $d$-wave superconductor quasiparticle scattering can
realize 2 or 4 copies of $\hh_{\mathsf{AIII}}^{\pup{2}}$ or $\hh_{\mathsf{AIII}}^{\pup{1}}$, respectively \cite{AltlandSimonsZirnbauer2002,Ghorashi2019}. 

The effective field theory describing the wave function and transport statistics
near zero energy
for the surface state models in Eqs.~(\ref{hAIII2}) and (\ref{hAIII1}) 
is a class AIII Wess-Zumino-Novikov-Witten (WZNW) nonlinear sigma model \cite{Nersesyan1994,Mudry1996,Caux1996,Bhaseen2001,Foster2014}
(see Sec.~\ref{sec:sigma}). 
It is identical in form to the principal chiral sigma model relevant to Gade-Wegner physics in class AIII \cite{GadeWegner1991,Gade1993,Guruswamy2000},
except that the latter lacks the WZNW term. 
The presence or absence of this term completely changes the low-energy physics \cite{Guruswamy2000,BernardLeClair2002,Evers2008}.
The key low-energy properties of Eqs.~(\ref{hAIII2}) and (\ref{hAIII1}) are as follows.

{\it{1.\ Zero-energy multifractal spectrum}}---Zero-energy 
eigenstates of $\hh_{\mathsf{AIII}}^{\pup{\nu}}$
(where $\nu \in\{1,2\}$ is the winding number)
are critically delocalized, with a purely parabolic multifractal
spectrum $\tau(q)$ \cite{Evers2008}. 
(For a review of multifractality in the context of topological 
superconductor surface states, see Refs.~\cite{Foster2014,Chou2014,Ghorashi2018,Ghorashi2019}.) 
The multifractal spectrum is \cite{Ludwig1994,Mudry1996,Caux1996,Chamon1996}
\begin{align}\label{tau(q)}
	\tau(q) =&\, 2(q - 1) + \Delta(q),
\nonumber\\
	\Delta(q) =&\, \theta_\nu \, q(1-q), \;\; 0 \leq |q| \leq q_c, \;\; q_c = \sqrt{\frac{2}{\theta_\nu}},
\end{align}
where the parameter $\theta_\nu$ is given by 
\begin{align}\label{theta_nu}
	\theta_\nu = \frac{\nu - 1}{\nu^2} + \frac{\lambda_A}{\pi}.
\end{align}
Here $\lambda_A$ is the variance of the abelian white noise disorder potential in Eqs.~(\ref{hAIII2}) and (\ref{hAIII1}),
\begin{align}\label{lamADef}
	\overline{A_{\bar{a},0}(\vex{r}) \, A_{\bar{b},0}(\vex{r}^\prime)}
	=
	\lambda_A
	\,
	\delta_{\bar{a},\bar{b}}
	\,
	\delta^\pup{2}(\vex{r}-\vex{r}^\prime).
\end{align}
The abelian disorder strength $\lambda_A$ is exactly marginal in the RG sense, 
i.e.\ it parameterizes a line of fixed points \cite{Ludwig1994,Mudry1996,Caux1996}.

In fact, Eq.~(\ref{tau(q)}) holds only for $q_c \geq 1$. For $q_c < 1$, the $\tau(q)$ 
undergoes ``freezing'' \cite{Chamon1996,Castillo1997,Carpentier2001,Chou2014}. 
A frozen wave function is quasilocalized, in that it typically exhibits a ``few''
rarified peaks. These peaks, however, are separated by arbitrarily large distances. 
We will only consider the unfrozen regime, i.e.\ $\lambda_A < \pi(2 - \frac{\nu - 1}{\nu^2})$ 
in this paper. 

The non-topological 2D Gade-Wegner models also feature an abelian parameter 
$\lambda_A$; this is related to the variance of the vector potentials $\{A_{\bar{a},0},A_{\bar{a},3}\}$ 
in the model defined by Eq.~(\ref{SDef}). 
Unlike the topological surface-state theory, however,
the $\lambda_A$ parameter always flows to infinity under the RG \cite{GadeWegner1991,Gade1993,Guruswamy2000}. 
For this reason, the low-energy physics of 2D Gade-Wegner models 
is always frozen \cite{Motrunich2002,Mudry2003}. 

{\it{2.\ Finite energy: correlation length and density of states}}---
A finite energy $\e \neq 0$ introduces a scale to the conformally invariant 
zero-energy theory. Formally, energy is a relevant coupling strength that flows to large
values under the renormalization group \cite{Ludwig1994,Ghorashi2018,Ghorashi2019}. We can associate to this 
relevant coupling a correlation length $\zeta(\e) \sim |E|^{-1/z}$, with the dynamical critical exponent \cite{Nersesyan1994,Mudry1996,Foster2014}
\begin{align}\label{z}
	z = 2 - \frac{1}{\nu^2} + \frac{\lambda_A}{\pi}. 
\end{align}
We will see that $\zeta(\e)$ plays an unconventional role
in the characterization of finite-energy eigenstates for topological class AIII
Hamiltonians [Eqs.~(\ref{hAIII2}) and (\ref{hAIII1})], in that it divides two different 
quantum critical scaling regimes, see Fig.~\ref{Fig--Summary} and Secs.~\ref{sec:nu=1}, \ref{sec:nu=2} and \ref{sec:lattice}. 
The length $\zeta(\e)$ thus plays the role of 
a \emph{crossover scale} between different critical fixed points, instead of dividing
a critical point (the delocalized zero-energy theory) from a massive fixed point
(which would be an Anderson insulator at finite energy).

The dynamical critical exponent $z$ also determines the scaling form of the low-energy density of states (DoS) $\rho(\e)$ 
for the Hamiltonians in Eqs.~(\ref{hAIII2}) and (\ref{hAIII1}), which
vanishes or diverges with a power-law, 
\begin{align}\label{DoS_nu}
	\lim_{\e \rightarrow 0} \rho(\e) \simeq |\e|^{(2 - z)/z}. 
\end{align}
Eq.~(\ref{DoS_nu}) assumes weak $\lambda_A$, such that the system is not ``frozen'' at 
zero energy (in the sense discussed above). 

Because $\lambda_A \rightarrow \infty$ in the Gade-Wegner models, the low-energy DoS 
in the 2D, non-topological version of class AIII \emph{always} freezes.
The low-energy behavior is given by 
\begin{align}\label{DoS_Gade}
	\rho_{\mathsf{Gade}}(\e)
	\sim 
	\frac{1}{|\e|}
	\exp\left(- c |\e|^{\alpha}\right)
\end{align}
where $\alpha = 1/2$ ($2/3$) below (above) freezing \cite{GadeWegner1991,Gade1993,Guruswamy2000,Motrunich2002,Mudry2003,DellAnnaFreezing2006}
(but see also \cite{Evers2014}). 

{\it{3.\ Zero-temperature, zero-energy conductivity}}---At
zero temperature and zero doping relative to the Dirac point,
the Landauer conductivity of $\hh_{\mathsf{AIII}}^{\pup{\nu}}$
is independent of the disorder \cite{Ludwig1994,Tsvelik1995,Ostrovsky2006}, 
and equivalent to that of the \emph{clean} limit \cite{Tworzydlo2006,Evers2008} 
\begin{align}\label{CondWZNW}
	\sigma_{x,x}
	=
	|\nu|/\pi,
\end{align}
in units such that the conductance quantum $e^2/h = 1$. 
This result also describes charge conduction through a wide, perfectly clean 
nanoscopic flake of graphene, precisely doped to the Dirac point \cite{Tworzydlo2006}.  
For the topological WZNW class AIII theory, 
Eq.~(\ref{CondWZNW}) appears to hold even in the presence of 
both disorder and interactions \cite{Xie2015}. 

By contrast, in the Gade-Wegner class AIII Dirac model,
the conductivity at zero energy is non-universal, depending 
on the microscopic strengths of the mass and potential disorders \cite{Guruswamy2000,Evers2008}. 
Interactions further modify the conductivity via Altshuler-Aronov corrections \cite{Foster2008,DellAnna2006};
the latter are excised by the WZNW term in the topological case \cite{Xie2015}.

\subsection{Integer quantum Hall plateau transition}

In Secs.~\ref{sec:nu=1} and \ref{sec:nu=2}, we will look for signatures of the 
class A quantum Hall plateau transition (QHPT) in the finite-energy eigenstates of 
Eqs.~(\ref{hAIII2}) and (\ref{hAIII1}). 
The key attributes we detect are the multifractal spectrum and Landauer transport properties. 
From numerical studies of the QHPT, the $\tau(q)$ spectrum is known to be 
approximately parabolic, given by Eq.~(\ref{tau(q)}) with $\theta_{\mathrm{QHPT}} \simeq 1/4$ \cite{Huckestein1995,Evers2008}. 
At the plateau transition, the disorder-averaged conductivity has been obtained from a Kubo-formula calculation 
of a disordered tight-binding model in a magnetic field \cite{Schweitzer2005}. 
Extrapolating results on very long samples with finite width to infinite width, the result is
\begin{align}\label{CondQHPT}
	\sigma_{x,x}^\pup{\mathrm{QHPT}}
	\simeq 
	0.58 \pm 0.02,
\end{align}
again in units such that $e^2/h = 1$. The distribution of the critical Landauer conductance $g$ 
has been computed for square samples of the Chalker-Coddington network model \cite{Jovanovic1998} 
and the tight-binding model \cite{Schweitzer2005}. 
In the limit of large system sizes, $g^\pup{\mathrm{QHPT}}=0.60 \pm 0.02$ was obtained \cite{Schweitzer2005}.
For earlier studies see Refs.\ \cite{Huckestein1995,Huo1993,Cho1997,Wang1998}.


\noindent
\begin{figure*}[t]
\noindent \begin{centering}
\includegraphics{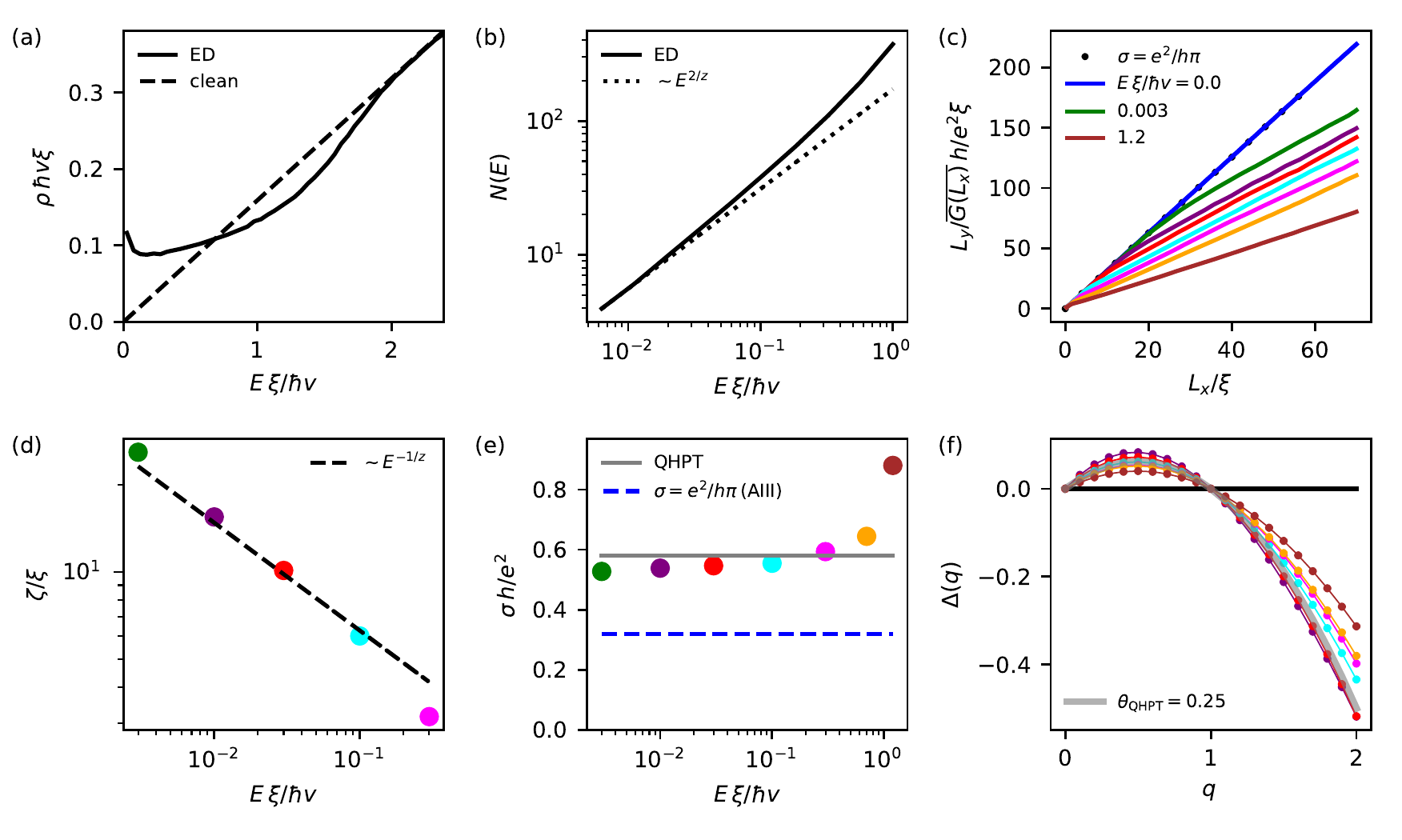}
\par\end{centering}
\caption{\label{fig:nu1}Numerical results for the 
topological class AIII surface model with a single Dirac node $(\nu=1)$, defined by Eq.~(\ref{hAIII1}) and protected by the 
``anomalous'' chiral symmetry in Eq.~(\ref{STopDef}).
The random vector potential strength is $W=2.3$ [Eq.~(\ref{eq:disCorr})]. 
(a) The DoS $\rho(E)$ versus energy, as calculated from ED (100 disorder realizations), is most strongly affected by disorder around the Dirac point ($E = 0$). 
(b) The integrated DoS $N(E) = \int_0^E d \epsilon \, \rho(\epsilon)$ is plotted versus energy. 
The predicted scaling form implied by Eq.~(\ref{DoS_nu}) is governed by the disorder-dependent dynamical critical exponent $z=1+W^{2}/\pi$. 
(c) Quantum transport results for the resistance normalized to system width, averaged over 200 disorder realizations. 
The energies are from top to bottom 
$E\xi/\hbar v=0,0.003,0.01,0.03,0.1,0.3,0.7,1.2$.
(d) The crossover scale from the transport calculation scales as the
correlation length $\zeta\sim E^{-1/z}$. 
(e) Conductivities extracted from the slope of the curves in panel (c) above the crossover scale $\zeta$, compared to the established
value of the QHPT critical conductivity. 
(f) Anomalous part of the multifractal spectrum $\Delta(q)$ extracted from box size scaling
of ED eigenstates for box sizes beyond the correlation length $\zeta$,
as determined in (d). The data correspond to 
$E\xi/\hbar v=0.01,0.03,0.1,0.3,0.7,1.2$
(bottom to top at $q=2$) and is based on 100 disorder realizations.}
\end{figure*}

\noindent
\begin{figure}
\noindent \begin{centering}
\includegraphics{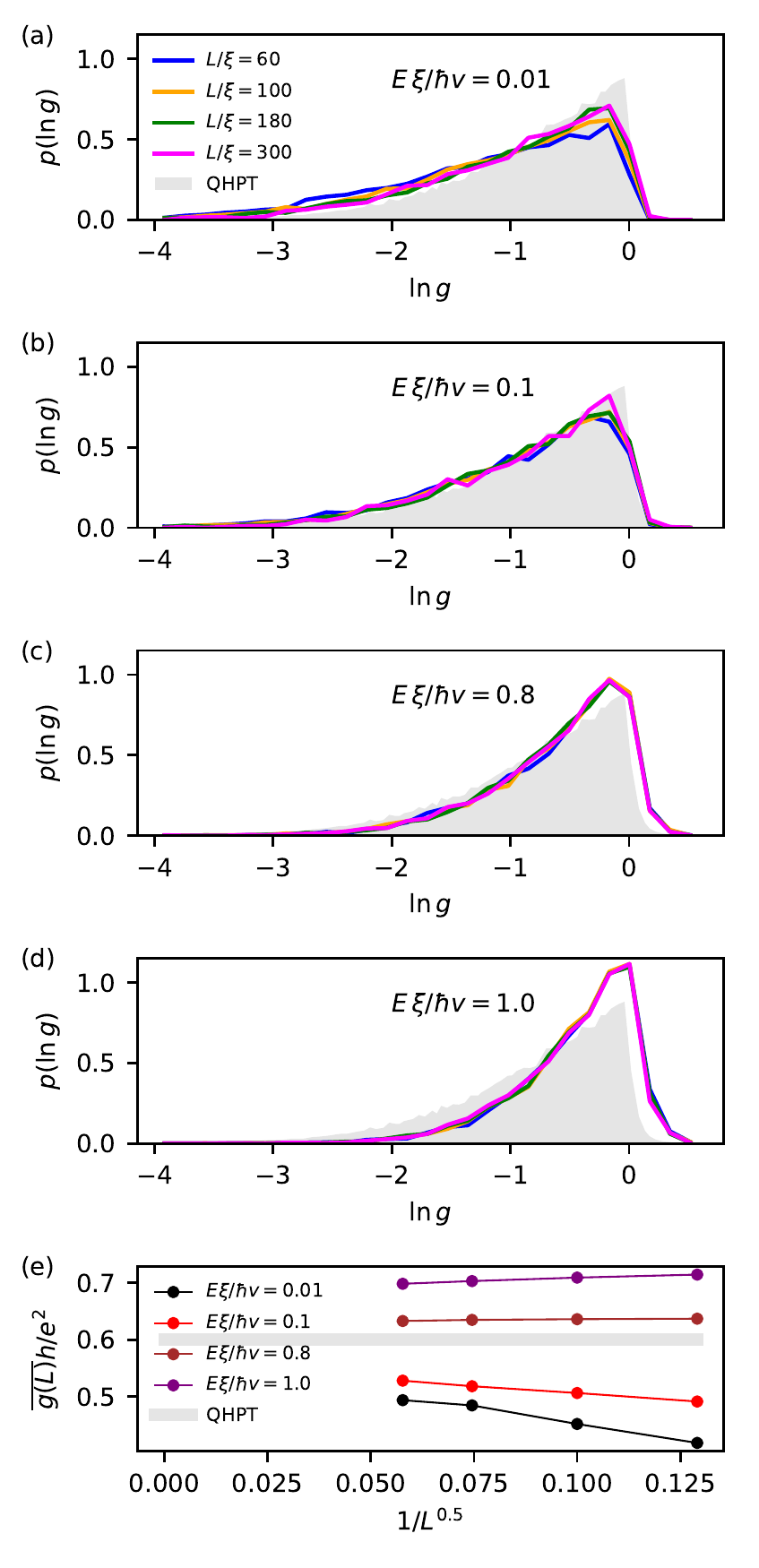}
\par\end{centering}
\caption{\label{fig:nu1-Pg}
Landauer conductance $g$ (in units of $e^2/h$) of square samples of length $L$ for the 
topological class AIII surface model with a single Dirac node $(\nu=1)$, as in Fig.~\ref{fig:nu1}. 
The probability distribution $p(\mathrm{ln}\,g)$ of the logarithm of the conductance for 
$L/\xi=60,100,180,300$ is depicted in panels (a)--(d) for energies $E\xi/\hbar v=0.01,0.1,0.8,1.0$, respectively. 
The data is based on between $1000$ and $10000$ disorder realizations, depending on system size. 
The grey filled area depicts the result obtained for a square-shaped Chalker-Coddington network model 
of size 128 reproduced from Ref.~\cite{Jovanovic1998}. Panel (e) shows the scaling of the 
mean square conductance $\overline{g(L)}$ for the above lengths and energies and the asymptotic 
value for the QHPT obtained from Ref.~\cite{Schweitzer2005}. 
For the horizontal axis, the exponent on $L$ has been chosen arbitrarily.  
}
\end{figure}

\section{One Dirac node, U(1) vector potential dirt (AIII, $\nu=1$)\label{sec:nu=1}}

In this section, 
we present numerical evidence for a continuous band of critical QHPT states in a single Dirac node with a random U(1) vector potential. 
This corresponds to the $\nu = 1$ class AIII surface state Hamiltonian in Eq.~(\ref{hAIII1}),
which is a $2 \times 2$ matrix differential operator, protected by the ``topological'' version of chiral symmetry in Eq.~(\ref{STopDef}).  
For analytical approaches, the white noise disorder correlator Eq.~(\ref{lamADef}) along with an UV cutoff on the Dirac dispersion is a convenient choice. 
In contrast, for our numerics, we find it useful to work with finite-range disorder correlations and take the UV-cutoff for the dispersion to infinity. 
We assume the disorder fields $A_{\bar{a},0}(\vex{r})$ ($\bar{a} \in \{1,2\}$)
to vary smoothly on a scale $\xi$, and these fields are taken to have 
zero average value over the sample area. The disorder statistics are taken to be Gaussian with
the correlator
\begin{equation}
	\overline{ A_{\bar{a},0}(\mathbf{r}) \, A_{\bar{b},0}(\mathbf{r}^{\prime}) }
	=
	\delta_{\bar{a},\bar{b}}
	\frac{W^{2}(\hbar v/\xi)^{2}}{2\pi}e^{-|\mathbf{r}-\mathbf{r}^{\prime}|^{2}/2\xi^{2}},
	\!\!
	\label{eq:disCorr}
\end{equation}
with disorder correlation length $\xi$; in this and the following two sections we restore $\hbar$.  
The dimensionless disorder
strength is 
$
	\frac{1}{\xi^{2}}
	\left(\frac{\xi}{\hbar v}\right)^{2}\int_{\mathbf{r}}\left\langle A_{1}(\mathbf{r}) \, A_{1}(\mathbf{r}^{\prime})\right\rangle _{\mathrm{dis}}
	=
	W^{2}
$.
In the limit of $\xi$ being the smallest scale, the exact analytical results \cite{Ludwig1994}
for the white noise disorder case should hold at low energies.
The dynamic critical exponent is predicted to be given by [Eq.~(\ref{z}) with $\nu = 1$]
\begin{align}\label{z1_Def}
	z = 1 + W^2/\pi.
\end{align}
This corresponds to the purely marginal parameter $\lambda_A = W^2$ in Eq.~(\ref{lamADef}),
which parameterizes a line of fixed points \cite{Ludwig1994}. 
These results are valid in the low-energy $|\e| \rightarrow 0$ limit, 
below the freezing transition that occurs at $W_{c}=\sqrt{2\pi} \simeq 2.5$ \cite{Chamon1996,Castillo1997,Carpentier2001,Chou2014}. 
In the following, we will work with $W=2.3$ in order to keep disorder-induced length
scales short while staying below $W_{c}$. We then expect $z = 2.68$.

We start with an 
exact diagonalization (ED) 
study of the Hamiltonian in Eq.~(\ref{hAIII1}), 
regularized on a lattice in momentum space, assuming periodic boundary conditions
in real space. Compared with tight-binding models in real space, the
momentum space approach avoids fermion doubling and band bending effects
on the one hand, but results in a dense Hamiltonian matrix which limits
the available system sizes on the other hand. We choose a linear dimension
of $L = 60 \xi$ and a momentum cutoff $\xi |k_{\bar{a}}| \leq R$ 
($\bar{a} \in \{1,2\}$)
with $R=5$ (corresponding to a matrix Hamiltonian of size 18050),
which gives sufficient real space resolution to resolve the smooth
variation of the disorder field. We checked that the results are converged
with respect to $R$ and the number of disorder realizations (typically
a few hundred).

In Fig.~\ref{fig:nu1}(a) we plot the ED-DoS $\rho$ versus energy
$E>0$. The scaling form $\rho(E)\sim E^{2/z-1}$ implies a divergence
at zero energy when $z>2$ as in our case. In our finite size system,
the divergence is replaced by a peak. At larger energies above $\sim2\hbar v/\xi$,
the DoS is unaffected by disorder scattering and asymptotically approaches
the clean value $\rho_{0}(E)=\frac{E}{2\pi\hbar v}$ (dashed line).
In Fig.~\ref{fig:nu1}(b) we quantify the DoS power law by plotting $N(E)=\int_{0}^{E}d\epsilon\,\rho(\epsilon)$
(solid) and confirm that it goes like $N(E)\sim E^{2/z}$ for small
$E$ (dashed line). 
This agreement with non-trival analytical predictions validates our numerical model 
and demonstrates precise control over the disorder strength.

We now turn to the results of a transport calculation for which we
assume $x$ to be the transport direction. The time independent Schr\"odinger
equation $\hh_{\mathsf{AIII}}^{\pup{1}} \, \psi = E \, \psi$ can be written as
\begin{align}
	\partial_{x}
	\psi\left(x,k_{y}\right) 
	=&\,
	\left(
		k_{y} \, \sigh_3 + i \sigh_1 \, k_{F}\right)
		\psi\left(x,k_{y}
	\right)
\nonumber\\
	&\,
	-
	i
	\hat{\sigma}_{1}
	\frac{1}{\hbar v}
	\sum_{k_{y}^{\prime}}
	\hat{U}(x,k_{y}-k_{y}^{\prime})
	\,
	\psi\left(x,k_{y}^{\prime}\right),
\end{align}
where 
$
	\hat{U}
	\equiv
	A_{1,0} \, \sigh_1 
	+
	A_{2,0} \, \sigh_2 
$ 
and $k_{F} \equiv E/\hbar v$.
This can be solved in terms of the transfer matrix, using the method
of Ref.~\cite{Bardarson2007}. We assume periodic boundary conditions
in the transverse direction with $|k_y| = 2 \pi |n|/L_y \leq R / \xi$ ($n \in \mathbb{Z}$) with $L_{y}=400\xi$,
and again use the mode cutoff
$R$ large enough so that the results below do
not depend on it. Assuming clean, highly doped leads attached at
$x = 0$ and $x = L_{x}$, the propagating lead modes 
are $\sigh_{1}$ eigenstates \cite{Bardarson2007}. This allows the definition of a
scattering matrix $S$. From the transmission block $t$, the longitudinal
conductance can be found as $G=\frac{e^{2}}{h}\mathrm{tr}\left(t^{\dagger}t\right)$.
We take the disorder average $\overline{G}$
and plot the resistance, normalized to the sample width $L_{y}/\overline{G}$
in Fig.~\ref{fig:nu1}(c) as a function of $L_{x}$. For a diffusive
sample, we expect $L_{y}/\overline{G}=L_{y}R_{0}+\frac{1}{\sigma}L_{x}$
where $R_{0}$ is some contact resistance and $\sigma$ the bulk conductivity. The data for $E=0$ agrees
with $\sigma=e^{2}/h\pi$ 
[Eq.~(\ref{CondWZNW})]
and $R_{0}=0$ for each disorder realization,
as required by gauge invariance and chiral symmetry \cite{Ostrovsky2006,Schuessler2009}.

For any finite energy, there is a crossover of the resistance trace
to diffusive behavior with a larger conductivity. We define the crossover
scale $\zeta$ as the length $L_{x}$ where $L_{y}/\overline{G}$
deviates by 5\% from the $E=0$ result.
In Fig.~\ref{fig:nu1}(d)
we show that the crossover scale for $E\ll\hbar v/\xi$ indeed follows
the scaling $\zeta\sim E^{-1/z}$ expected for the correlation length.

In Fig.~\ref{fig:nu1}(e), we plot the conductivities as fitted from
the resistance data well above the crossover scale for $L_x>\zeta+10\xi$. 
As the resistance trace is not perfectly linear in $L_x$, there is a few percent 
ambiguity in the definition of $\sigma$ that we further address in the next paragraph.
For $0<E\lesssim\hbar v/\xi$,
the conductivities show a plateau at 
$
	\sigma 
	\simeq 
	0.55
	\frac{e^{2}}{h}
$
in fair agreement with the value 
$
	\sigma_{x,x}^\pup{\mathrm{QHPT}}
	=
	0.58 \pm 0.02 \frac{e^{2}}{h}
$
obtained by Schweitzer and Marko\v s \cite{Schweitzer2005} via the
Kubo formula for a lattice model at the QHPT transition. At larger
energies $E$, the conductivity at the accessible length scales increases
with energy.
This is as expected for the 
crossover to the
semiclassical Drude conductivity,
which goes as $\sigma \sim (e^2/h)(1/W^2)$ \cite{FosterAleiner2008}. 

To get further insight into the behavior of quantum transport as a function of system size, 
we next study the probability distribution of the Landauer conductance $g(L)$ for square samples 
with increasing length $L$, keeping the disorder strength the same as above. The results for energies 
$E\xi/\hbar v=0.01,0.1,0.8,1.0$ 
are shown in Fig.~\ref{fig:nu1-Pg}, panels (a)--(d). In all cases, the distributions $p(\mathrm{ln}\,g)$ 
agree reasonably well with the established distribution obtained from a square shaped Chalker-Coddington network 
model of size 128, reproduced from Ref.~\cite{Jovanovic1998} (grey shaded area). For the energies 
$E\xi/\hbar v = 0.01,0.1$, an evolution of the distribution towards the Chalker-Coddington result can be 
observed with increased system size $L$, 
whereas for $E\xi/\hbar v=0.8,1.0$, 
the distribution does not change visibly on the available length scales. 
The evolution of the average square conductance $\overline{g(L)}$ with $L$ is depicted in Fig.~\ref{fig:nu1-Pg}(e). 
For energies $E\xi/\hbar v=0.01,0.1$, we find an increase with system size, whereas there is a weak decrease 
for $E\xi/\hbar v=0.8,1.0$ (not discernible in the conductance distributions). 
The data for all energies is consistent with a limiting value of $g^\pup{\mathrm{QHPT}}=0.60\pm0.02$, 
as obtained in Ref.~\cite{Schweitzer2005} by extrapolating results for a disordered lattice model 
to large system sizes (grey line). 
For the horizontal axis in panel (e), the exponent $-0.5$ on $L$ 
has been chosen arbitrarily. 
Our range of length values is not sufficient to determine if there is indeed a power law approach 
of $\overline{g(L)}$ to its limiting value with a characteristic irrelevant exponent, 
as established for the fine-tuned QHPT in Ref.~\cite{Schweitzer2005}. 
Our data suggest that the relevant length scale for the approach to the QHPT fixed point 
strongly increases with energy beyond $E\simeq \hbar v/\xi$. 
Interestingly, this parallels the behavior of the Boltzmann transport mean free path 
for our correlated disorder model where the scattering range in momentum space is 
restricted to $1/\xi$.

Further evidence for the presence of a stack of QHPT critical states
at finite energies comes from the anomalous part of the multifractal
spectrum $\Delta(q) = \tau(q)-2(q-1)$, as shown in Fig.~\ref{fig:nu1}(f).
Here, $\tau(q)$ is defined via the scaling of the disorder averaged
generalized inverse participation ratios \cite{Evers2008},
\begin{equation}
	\overline{ P_{q} }
	\sim
	(b/L)^{\tau(q)},
	\label{eq:tauDef}
\end{equation}
where, for a single disorder realization,
\begin{align}
	P_{q}
	=
	\sum_{\mathcal{B}}
	\left(
		\sum_{\mathbf{r}_{i}\in\mathcal{B}}
		|\psi(\mathbf{r}_{i})|^{2}
	\right)^{q}
\end{align}
and the outer sum is over square regions $\mathcal{B}$ with linear
size $b$ covering the real space lattice. To find $P_{q}$, we use the ED eigenstate $\psi$ closest to energy $E$. It turns out that for small energies
$
	E
	\apprle
	0.1 \hbar v/\xi,
$ 
the power law in Eq. (\ref{eq:tauDef})
is governed by two different multifractal spectrums 
$\tau(q)$, with the crossover at 
$b \simeq \zeta$
as found in Fig.~\ref{fig:nu1}(d). To capture only the long distance physics
beyond the correlation length, we restrict to  
$b > \zeta$,
which is only possible for $E \geq 0.01 \hbar v/\xi$ due to the overall system size restriction. For a wide range of energies
$E \apprle \hbar v/\xi$, we find $\Delta(q)$ in good agreement with
the established form for QHPT states~\cite{Huckestein1995,Evers2008}, 
$\Delta(q) = \theta_{\mathrm{QHPT}} \, q(1-q)$ [Eq.~(\ref{tau(q)})],
with $\theta_{\mathrm{QHPT}} \simeq 0.25$. 
The class AIII multifractality from Eq.~(\ref{theta_nu}) expected at low energies and 
for box sizes below the length scale $\zeta$ is not clearly visible in our numerics. 
This is due to insufficient system size for the high degree of rarification associated 
to our large disorder strength (chosen just below the freezing transition \cite{Chamon1996,Castillo1997,Carpentier2001}). 
For a similar study at weaker disorder strength confirming Eq.~(\ref{theta_nu}) 
at low energies, see Ref.~\cite{Chou2014}.


\section{Two Dirac nodes, U(1) $\oplus$ SU(2) vector potential dirt (AIII, $\nu=2$)\label{sec:nu=2}}

\noindent
\begin{figure*}[t!]
\noindent \begin{centering}
\includegraphics{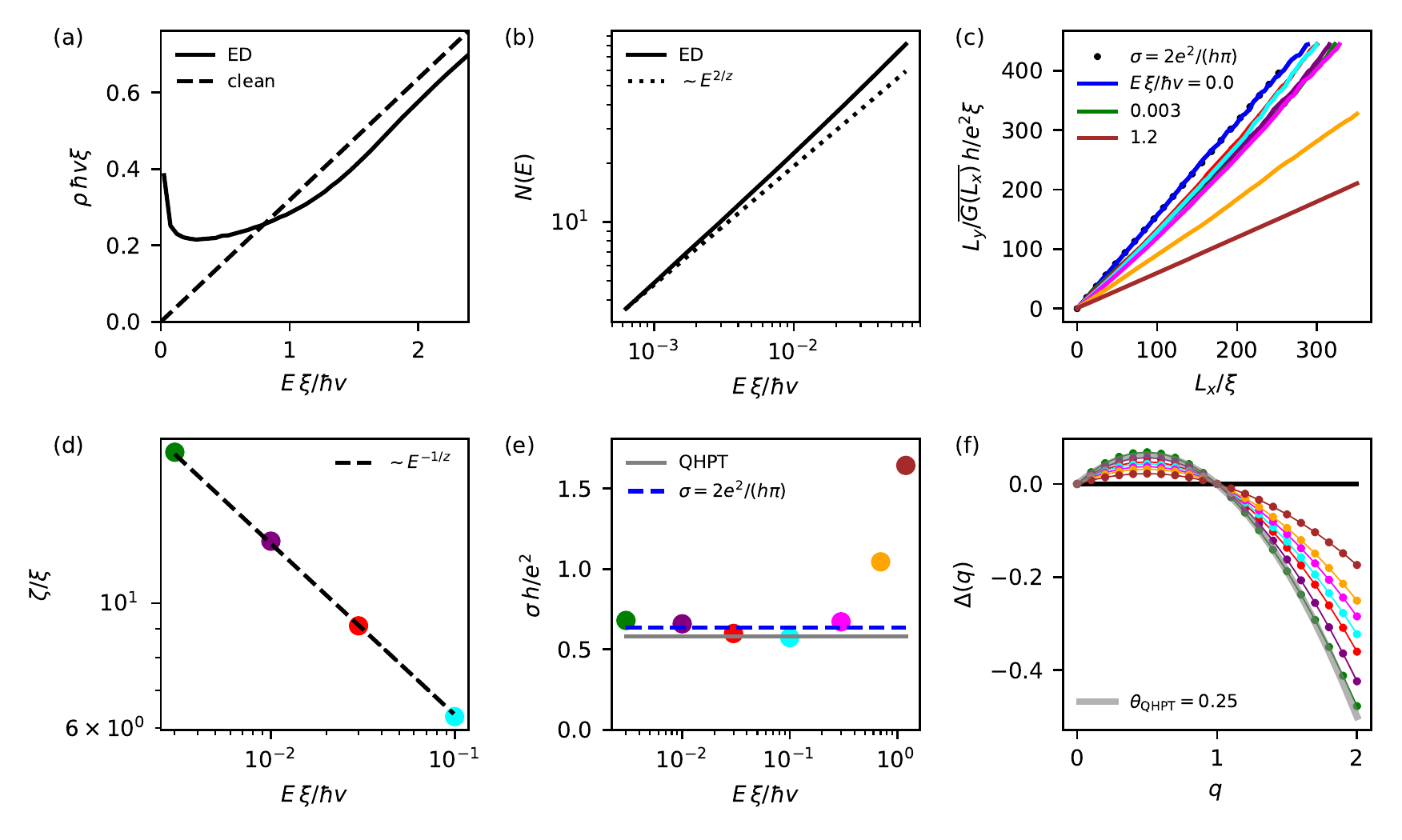}
\par\end{centering}
\caption{\label{fig:nu2}Numerical results for the 
topological class AIII surface model with two Dirac nodes $(\nu=2)$, defined by Eq.~(\ref{hAIII2}) and protected by the 
``anomalous'' chiral symmetry in Eq.~(\ref{STopDef}).
The Abelian and non-Abelian vector potential disorder strengths are 
$W_{A}=2.2$ and $W_{N}=1.5$, respectively [\emph{cf.}\ Eq.~(\ref{eq:disCorr})]. 
(a) The DoS as a function of energy, as calculated from ED (200 disorder realizations). 
(b) The integrated DoS $N(E) = \int_0^E d \varepsilon \, \rho(\varepsilon)$ plotted versus energy. 
The predicted scaling form implied by Eq.~(\ref{DoS_nu}) is governed by the dynamical critical exponent $z = 7/4+W_{A}^{2}/\pi$,
which depends only on the Abelian disorder strength. 
(c) Quantum transport results for the resistance normalized to system width $L_y=400\xi$, averaged over 600 disorder realizations. 
The energies are from
top to bottom $E\xi/\hbar v=0,0.003,0.01,0.03,0.1,0.3,0.7,1.2$. 
(d)
The crossover length from the transport calculation scales as the
correlation length $\zeta\sim E^{-1/z}$. 
(e) Conductivities extracted
from the $L_{x}\protect\geq200\xi$ slopes of the curves in panel
(c), compared to the established value of the QHPT critical conductivity.
(f) Anomalous part of the multifractal spectrum $\Delta(q)$ extracted
from box size scaling of ED eigenstates for box sizes beyond the correlation
length $\zeta$ as extracted in (d). 
The data correspond to $E\xi/\hbar v=0.003,0.01,0.03,0.1,0.3,0.7,1.2$
(bottom to top at $q=2$). 
}
\end{figure*}

\noindent
\begin{figure}
\noindent \begin{centering}
\includegraphics{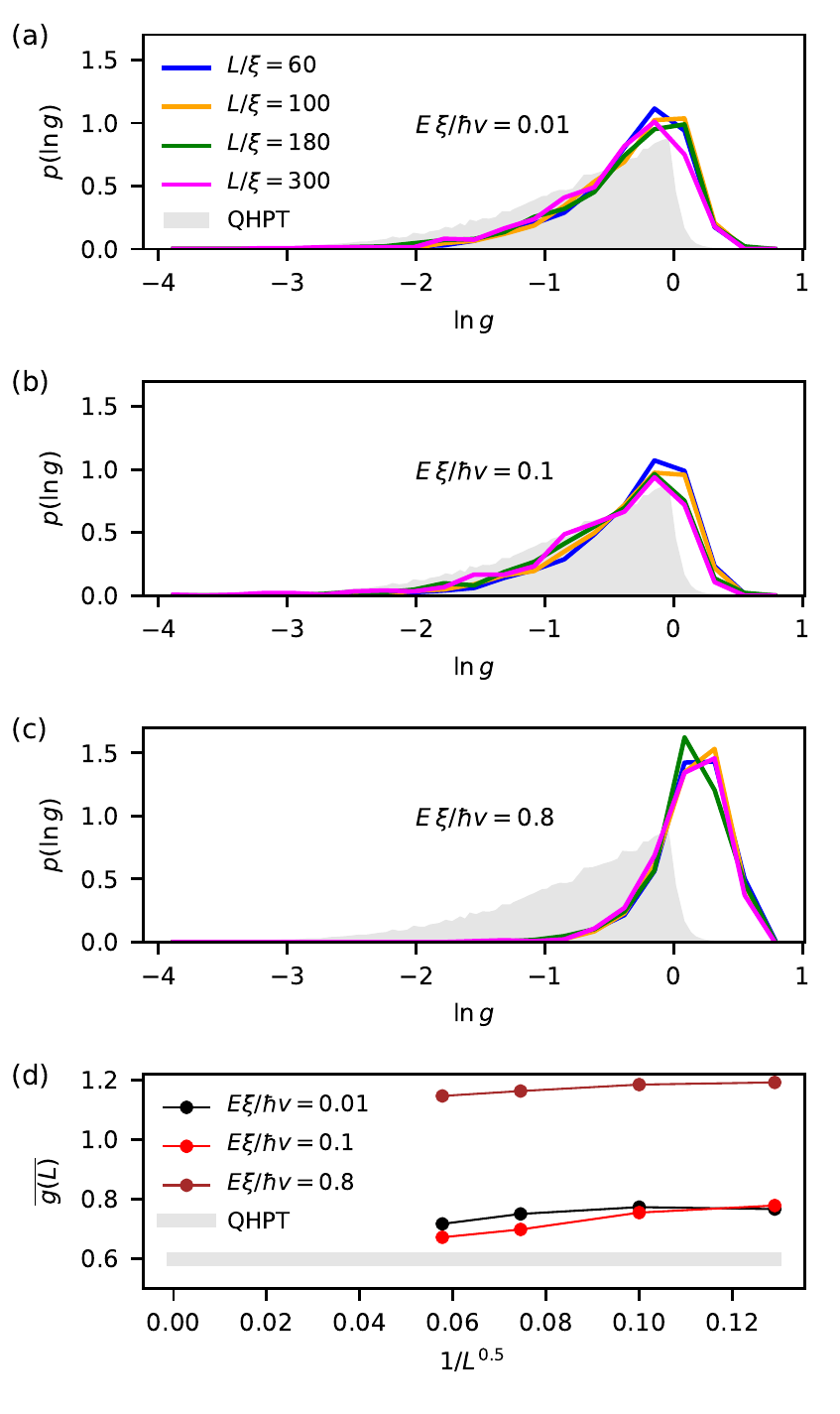}
\par\end{centering}
\caption{\label{fig:nu2-Pg}
Landauer conductance $g$ (in units of $e^2/h$) of square samples of length $L$ for the 
topological class AIII surface model with two Dirac nodes $(\nu=2)$, as in Fig.~\ref{fig:nu2}. 
The probability distribution $p(\mathrm{ln}\,g)$ of the logarithm of the conductance for $L/\xi=60,100,180,300$ 
is depicted in panels (a)--(c) for energies $E\xi/\hbar v=0.01,0.1,0.8$, respectively. The data is 
based on between $3000$ and $600$ disorder realizations, depending on system size. The grey filled area depicts 
the result obtained for a square shaped Chalker-Coddington network model of size 128, reproduced from 
Ref.~\cite{Jovanovic1998}. Panel (d) shows the scaling of the mean square conductance $\overline{g(L)}$ 
for the above lengths and energies and the asymptotic value for the QHPT obtained from Ref.~\cite{Schweitzer2005}. 
For the horizontal axis, the exponent on $L$ has been chosen arbitrarily. 
}
\end{figure}

\begin{figure}
\includegraphics{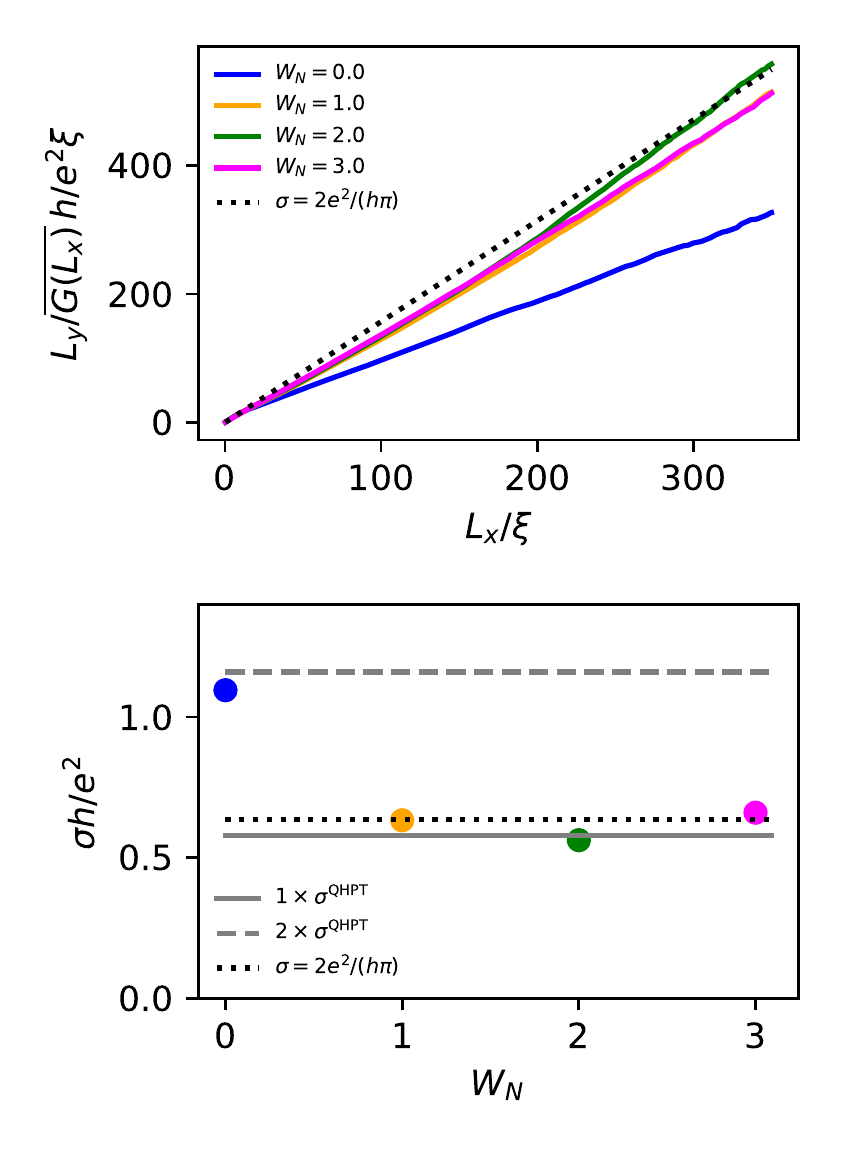}
\caption{\label{fig:nu2-WNsweep}Numerical transport results 
for the topological two-node class AIII Dirac model [$\nu = 2$, Eq.~(\ref{hAIII2})],
with Abelian and non-Abelian vector potential disorder
of strengths $W_{A}=2.1$ and increasing $W_{N}=0,1,2,3$ at energy
$E=0.03\hbar v/\xi$. The top panel shows the bare resistance data,
while the lower panel depicts the bulk conductivities obtained from
linear fits to the bare resistance data above $L_{x}=200\xi$.
These plots establish the crossover of the two-node model from the
finite-energy conductivity plateau equal to 
$2\times\sigma_{x,x}^\pup{\mathrm{QHPT}}$ in the absence of internode scattering,
to a plateau with value 
$1\times\sigma_{x,x}^\pup{\mathrm{QHPT}}$ in its presence. The data is obtained for systems of width $L_y=400\xi$ and is averaged over 600 disorder realizations.
}
\end{figure}

\noindent
\begin{figure*}[t!]
\noindent \begin{centering}
\includegraphics{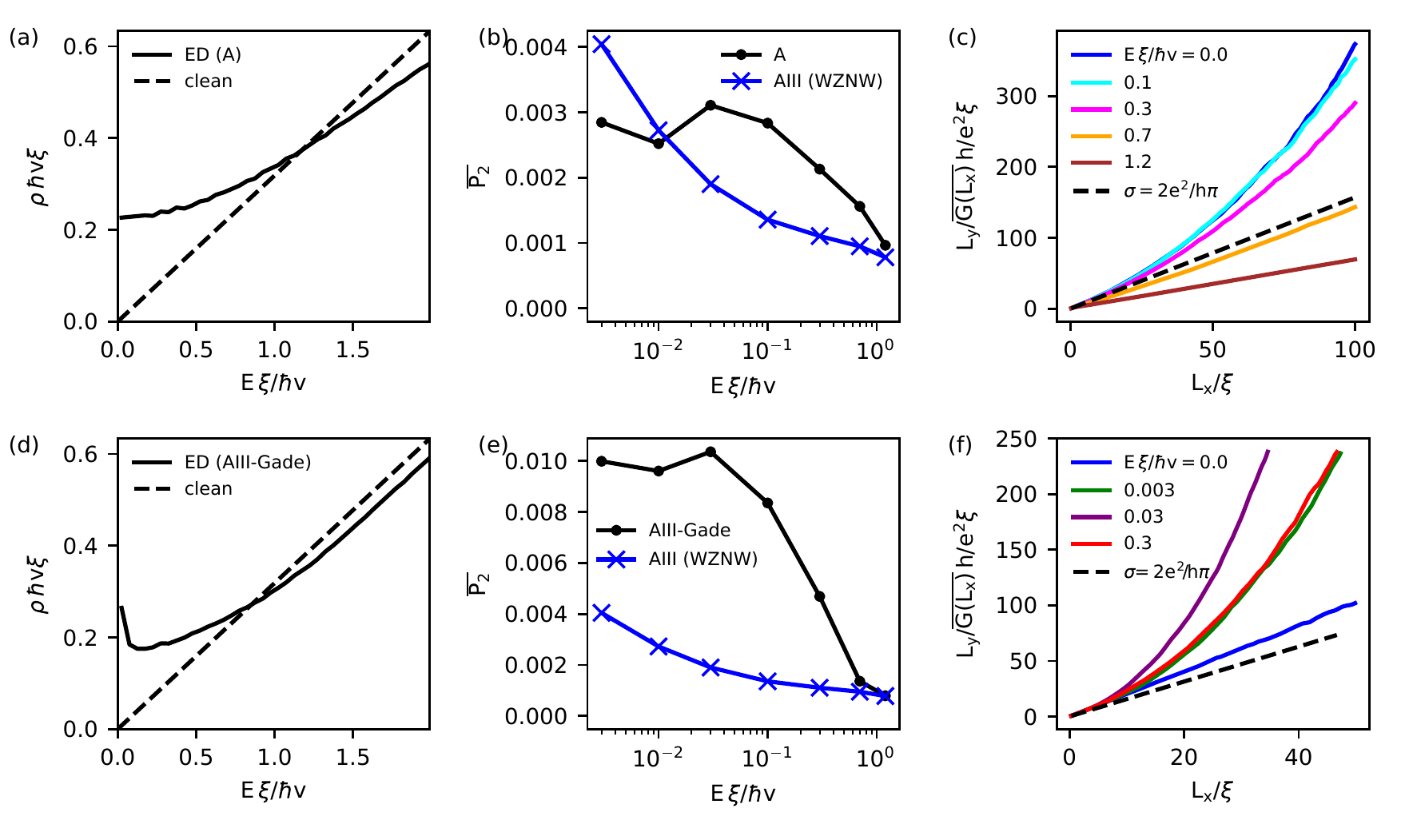}
\par\end{centering}
\caption{\label{fig:nu2-localizing}
Numerical results for non-topological, two-node Dirac models. 
Panels (a,b,c) show results for the unitary class A model defined by all 16 disorder potentials
in Eq.~(\ref{hDef}). 
Panels (d,e,f) show results for the Gade-Dirac class AIII model defined by restricting 
to the 8 potentials allowed by the sublattice symmetry in Eq.~(\ref{SDef}). 
In both cases, all disorder strengths are set equal to $W=1.5$. 
The DoS versus energy is shown in panels (a) and (d) respectively,
(b) and (e) depict the IPR and compare to the results for the topological class AIII
model from Sec.~\ref{sec:nu=2} and Fig.~\ref{fig:nu2}. The data is averaged over 100 disorder realizations. Quantum transport results
showing the resistance versus sample length are depicted in panels
(c) and (f). The data is averaged over 200 disorder realizations. }
\end{figure*}

In this section, we consider the $\nu = 2$ class AIII topological surface
state model in Eq.~(\ref{hAIII2}).
The disorder consists of U(1) $\{A_{\bar{a},0}\}$ and SU(2) $\{A_{\bar{a},i} \, \tauh^i\}$
vector potentials, with $\bar{a} \in \{1,2\}$ and $i \in \{1,2,3\}$.  
We take these eight potentials to be mutually uncorrelated,
each with a smooth autocorrelation function as in Eq.~(\ref{eq:disCorr}). 
The disorder strength is taken to be $W_{A}$ for each of the two Abelian
potentials and $W_{N}$ for the non-Abelian ones, respectively. We
set $W_{N}=1.5$ and $W_{A}=2.2$ to stay below the freezing transition
at $W_{A,c}=\sqrt{{7\pi}/{4}\pi}=2.34$ [freezing is discussed below Eq.~(\ref{lamADef})].  
We obtain $z=7/4+W_{A}^{2}/\pi=3.29$ for the dynamical critical exponent [Eq.~(\ref{z})].

As above for $\nu=1$, we start with an ED study of Hamiltonian $\hh_{\mathsf{AIII}}^{\pup{2}}$
and use $L=60 \xi$ and $R=5$ as a UV-cutoff parameter (corresponding to a matrix Hamiltonian of size 36100). 
Fig.~\ref{fig:nu2}(a)
shows the DoS $\rho(E)$ with the expected peak at energy $E=0$ and
a crossover to the clean DoS of two Dirac cones (dashed line) at large
energies. In Fig.~\ref{fig:nu2}(b) we confirm the expected power
law of the DoS $\rho(E)\sim E^{2/z-1}$ by plotting $N(E)\sim E^{2/z}$.

We now turn to quantum transport results, obtained as in the $\nu=1$
case. We use samples of width $L_{y}=400\xi$ and increase the length
$L_{x}$ up to $350\xi$. Fig.~\ref{fig:nu2}(c) shows the resistance
normalized to sample width as a function of length $L_{x}$ for a
range of energies between $E=0$ and $E=1.2\hbar v/\xi$. We define
a crossover length $\zeta$ from the resistance data as above for
$\nu=1$ and confirm the expected scaling $\zeta\sim E^{-1/z}$ in
Fig.~\ref{fig:nu2}(d) for energies $E\lesssim0.3\hbar v/\xi$. Beyond
this crossover length, the slope of the low energy resistance curves
in Fig.~\ref{fig:nu2}(c) increases and then roughly saturates for $L_{x}\gtrsim200\xi$.
The conductivity fitted from this large $L_{x}$ data is plotted versus
energy in Fig.~\ref{fig:nu2}(e); we find values in close vicinity
to 
$\sigma_{x,x}^\pup{\mathrm{QHPT}}$ [Eq.~(\ref{CondQHPT})]
for small $E$ and conductivities increasing with
energy for $E\apprge0.3\hbar v/\xi$.
To further investigate the scaling of transport properties with system size, 
we study the probability distribution of the Landauer conductance $g(L)$ of square samples of length $L$, 
keeping the disorder strength the same as above. The results for energies $E\xi/\hbar v=0.01,0.1,0.8$ are 
shown in Fig.~\ref{fig:nu2-Pg}, panels (a)--(c), 
and 
the evolution of the average conductance $\overline{g(L)}$ with $L$ is depicted in Fig.~\ref{fig:nu2-Pg}(d). 
The situation is similar to the $\nu=1$ case. While small energies $E\xi/\hbar v=0.01,0.1$ result in distributions 
and average square conductances visibly approaching the QHPT case, at higher energy $E\xi/\hbar v=0.8$ the relevant 
length scale for this process exceeds our available system size and only a slight trend of the average conductance is visible.

Finally, Fig.~\ref{fig:nu2}(f) shows the anomalous part of the multifractal
spectrum, as calculated from the ED eigenstates in the vicinity of
the energies used for the transport study. In analogy with the conductivities,
we find good agreement with the QHPT form for low energies. At higher
energies weaker multifractality is observed.

Before closing this section, we give a complementary perspective on
the above results by ramping up from zero the non-Abelian disorder strength. We
take $W_{N}=0,1,2,3$ at fixed Abelian disorder strength $W_{A}=2.1$
and fixed energy $E=0.03\hbar v/\xi$ and present transport data in
Fig.~\ref{fig:nu2-WNsweep}. For $W_{N}=0$, the two nodes are decoupled
and we find a conductivity close to $2\times\sigma_{x,x}^\pup{\mathrm{QHPT}}$ in agreement
with the single node case presented in Sec.~\ref{sec:nu=1}.
For $W_{N}=1,2,3$ the nodes are coupled and the conductivity is close
to the value $1\times\sigma_{x,x}^\pup{\mathrm{QHPT}}$.


\section{Non-topological two-Dirac node models: Gade-Wegner AIII, localized graphene \label{sec:nontop}}

We now contrast our findings for 
the topological-surface-state class AIII models presented in Secs.~\ref{sec:nu=1} and \ref{sec:nu=2}
with two non-topological Dirac models, both of which are expected to Anderson localize at finite energies. 

We first consider the generic 2-node Dirac (``spinless graphene'') model defined by $\hh \equiv \hh_0 + \hh_A + \hh_B + \hh_C$ 
[Eq.~(\ref{hDef})]. This model possesses only U(1) symmetry, without $T$, $P$, $S$ [Eq.~(\ref{Latt_Sym})] or the topological $\Stop$ [Eq.~(\ref{STopDef})] symmetries.
All disorder potentials are taken to be short-range correlated as in Eq.~(\ref{eq:disCorr}), with zero average. 
This model resides in the unitary class with a vanishing Hall conductivity, and is expected to localize at all energies \cite{Ostrovsky2006,Evers2008}. 
All disorder potentials are taken to carry the same strength $W=1.5$. 
In Fig.~\ref{fig:nu2-localizing}, we show
numerical results for this model in the top row, panels (a,b,c). The
numerical results are obtained analogously to those presented in the previous sections.
The DoS in panel (a) is finite and featureless at and around zero
energy. In panel (b), we show the IPR $\overline{ P_{q=2} }$,
where we use a box size corresponding to the disorder correlation
length $b\simeq\xi$. While the IPR is of the same order as for class
AIII-WZWN (crosses), the localizing behavior is clearly observed in
the resistance data in panel (c) probing larger length scales. For
small energies $E\leq0.3\hbar v/\xi$ the resistance is growing with
increasing slope to values much above the clean zero energy resistance
(dashed line).

We also consider the non-topological Gade-Wegner class AIII Dirac model,
defined by imposing the sublattice symmetry in Eq.~(\ref{SDef}). 
Just as in the topological case studied in Sec.~\ref{sec:nu=2}, there are eight allowed
disorder perturbations. The crucial difference from the topological model in Eq.~(\ref{hAIII2})
is that Eq.~(\ref{SDef}) permits mass and scalar potentials, in addition to vector potential disorder.
The allowed perturbations from Eq.~(\ref{hDef}) are $\{m_1,m_2,A_{1,3},A_{2,3},V_1,V_2,A_{1,0},A_{2,0}\}$ \cite{Guruswamy2000}. 
Numerical results are shown in Fig.~\ref{fig:nu2-localizing}(d,e,f). The DoS in panel
(d) shows the characteristic Gade-type singularity at $E=0$ [Eq.~(\ref{DoS_Gade})],  
and the IPR in panel (e) is increased as compared to the topological version from Fig.~\ref{fig:nu2}. 
Concurrently, the tendency to localization in the resistance plot in (f) is clearly
observable. However, note that the conductivity at zero energy in
the Gade class AIII Dirac model is finite and slightly lower than the clean conductivity
$\sigma=2e^{2}/(h\pi)$ (dashed line), in agreement with theoretical predictions \cite{Guruswamy2000,Ostrovsky2006}.


\begin{figure}
\centering
\includegraphics[width=0.23\textwidth]{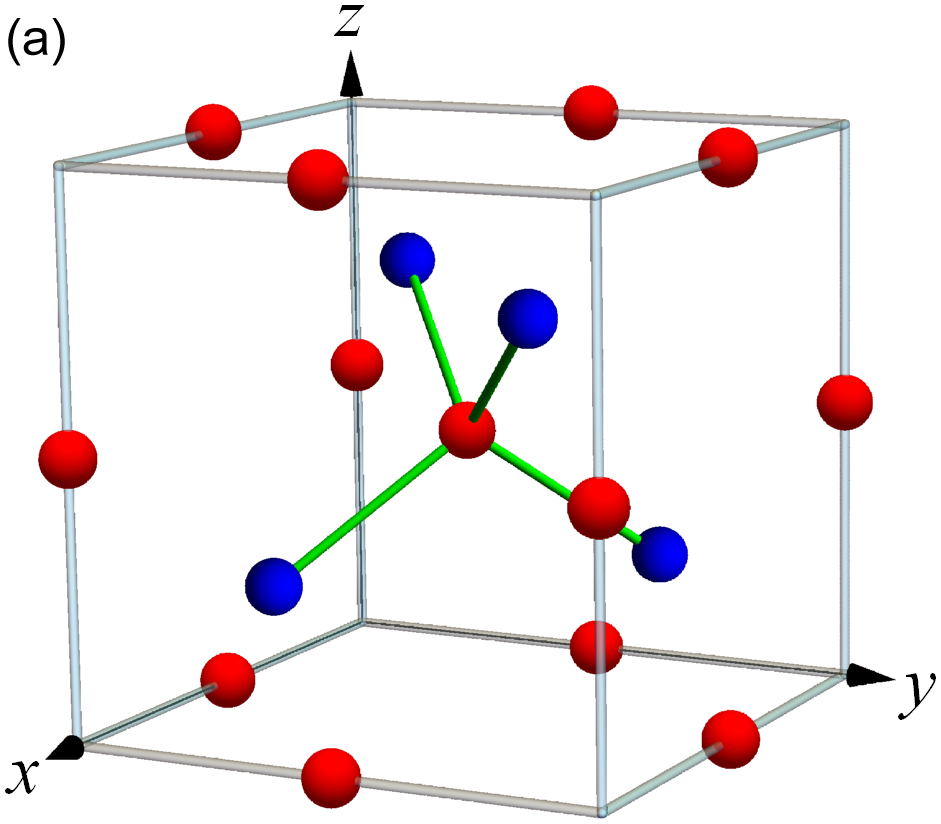}
\includegraphics[width=0.23\textwidth]{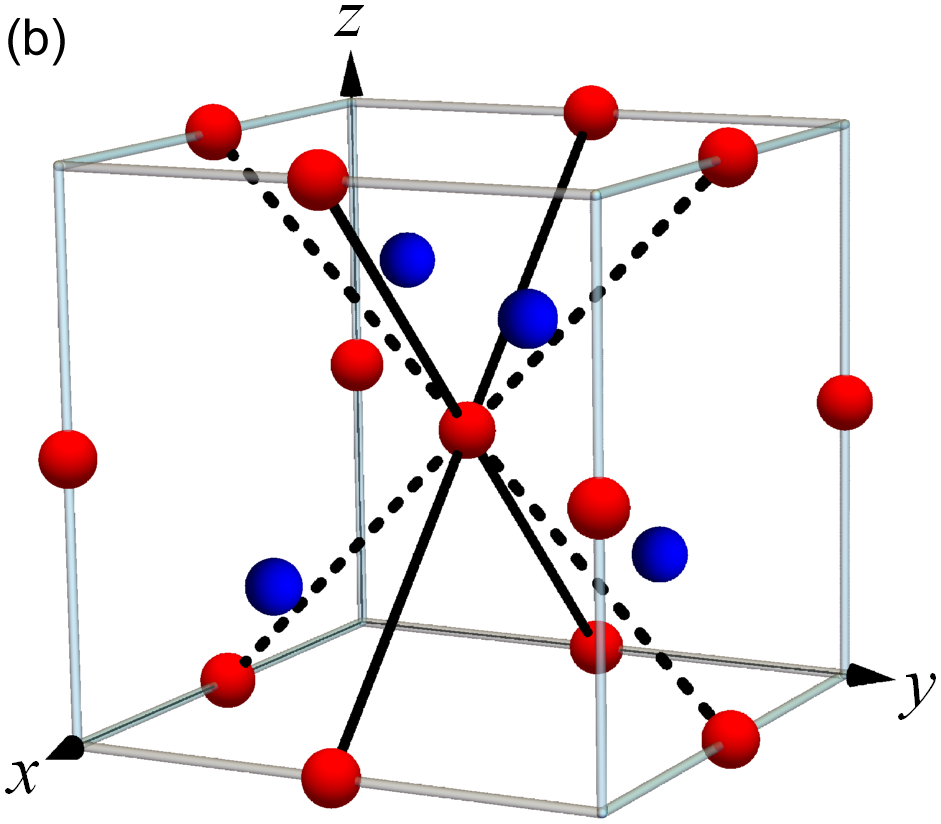}\\
\includegraphics[width=0.23\textwidth]{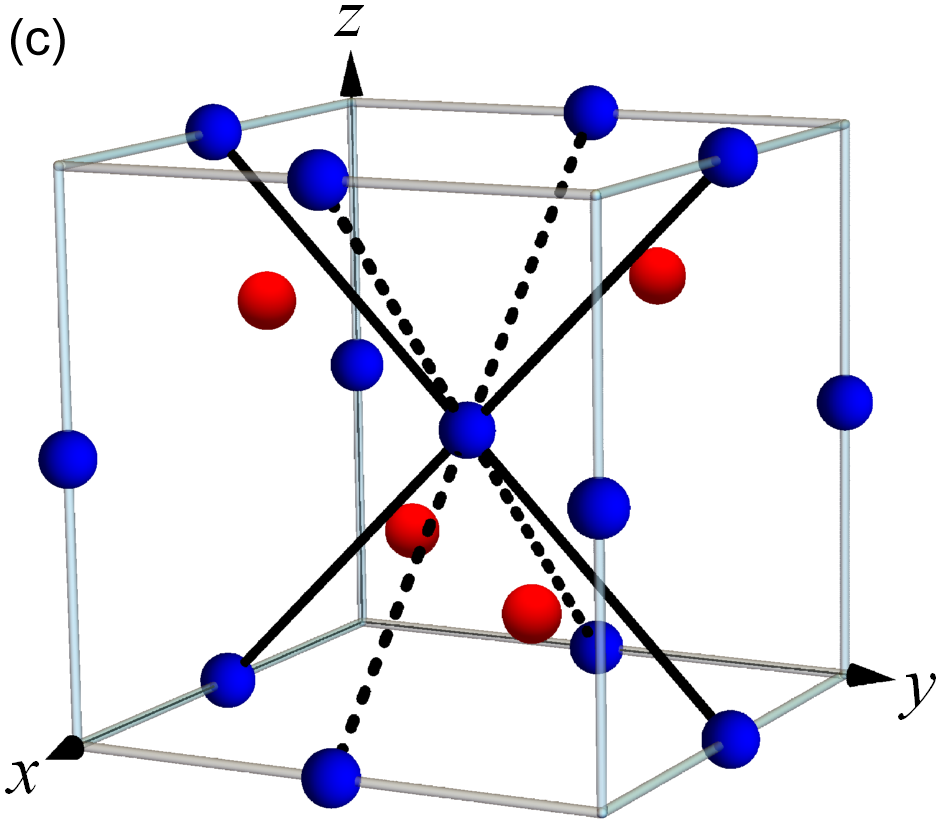}
\includegraphics[width=0.23\textwidth]{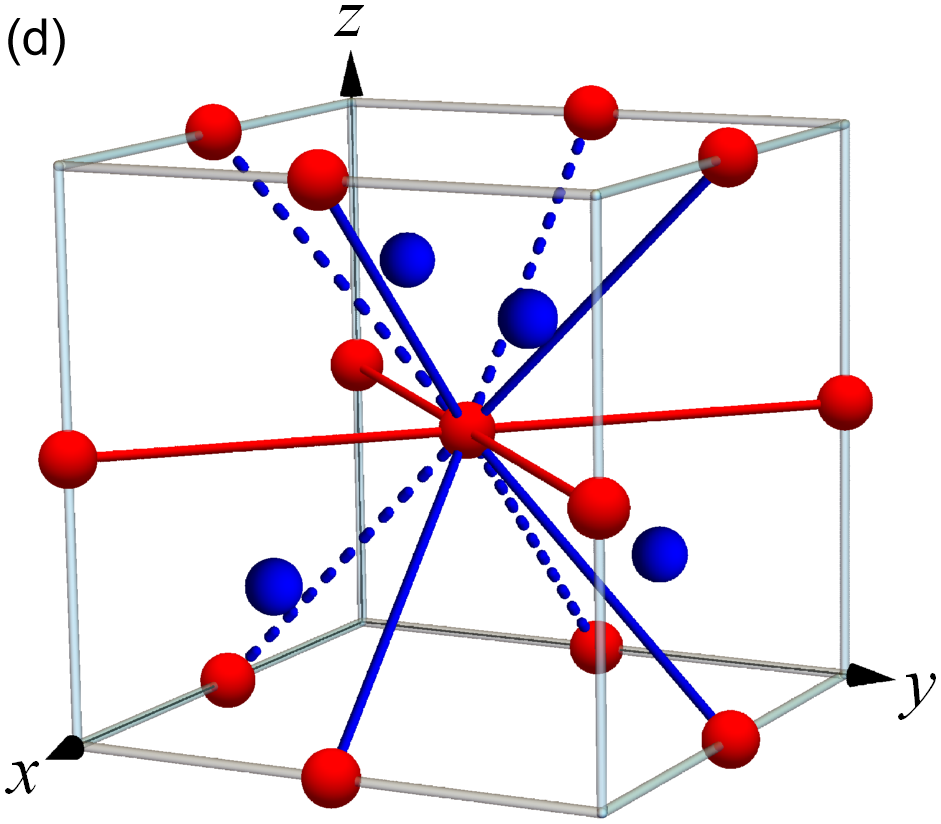}\\
\caption{Class AIII topological superconductor
model on the diamond lattice \cite{Xie2015}.
Sublattice $A$ ($B$) sites are indicated by red (blue) spheres, and 
are subject to the on-site potential $\mu_s$ ($-\mu_s$) [Eq.~(\ref{ViDef})].
(a) Nearest-neighbor hopping with amplitude $t^\prime$ [Eq.~(\ref{H1})]. 
(b) Next-nearest-neighbor hopping on the $A$ sublattice [Eqs.~(\ref{H2}) and (\ref{tiDef})]. 
Solid (dashed) hopping amplitudes are equal to $t/2$ (-$t/2$). 
(c) Next-nearest-neighbor hopping on the $B$ sublattice [Eqs.~(\ref{H2}) and (\ref{tiDef})]. 
Solid (dashed) hopping amplitudes are equal to $t/2$ (-$t/2$). 
(d) BCS pairing within each of the sublattices [Eqs.~(\ref{H3}) and (\ref{Deltas})]. 
The red lines indicate $d$-wave spin-singlet pairing in the $xy$-plane,
i.e.\ red bonds create singlet Cooper pairs with amplitude $\Delta/2$. 
The blue lines indicate $p$-wave $z$-axial spin-triplet pairing in the 
$xz$- and $yz$-planes. 
Solid (dashed) bonds create total spin $J^z = 0$ pairs with amplitude 
$\Delta/2$ ($-\Delta/2$).} 
\label{fig:diamond-lattice}
\end{figure}

\begin{figure}[b!]
	\centering
	\includegraphics[width=.6\columnwidth]{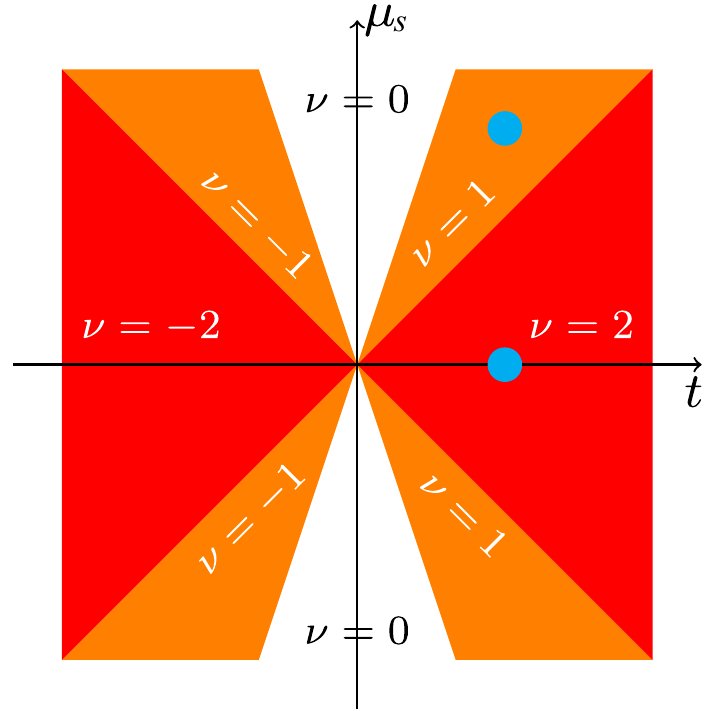}
	\caption{Phase diagram of the lattice model defined by 
	Eqs.~(\ref{H1})--(\ref{Deltas}),
	as function of 
	the staggered chemical potential $\mu_s$ 
	and the next-nearest-neighbor hopping $t$ [Eqs.~(\ref{H2})--(\ref{tiDef})]. 
	There are critical lines at 
	$\mu_s = \pm t$ 
	and 
	$\mu_s = \pm 3t$ 
	separating 
	different topological phases with winding numbers $\nu=-2,-1,0,1,2$. 
	For $\nu = 2$, the bulk gap is largest 
	(and therefore the surface penetration depth smallest) 
	at $\mu_s = 0$ and finite $t>0$. 
	The $\nu=1$ phase is narrower in phase space and harder to observe in 
	ED numerics due to the larger penetration depth.
	The blue dots indicate the $\nu = 2$ and $\nu = 1$
	parameters investigated in Figs.~\ref{fig:lattice_result} and \ref{fig:lattice_spectra}.
	}
	\label{fig:lattice_phases}
\end{figure}

\noindent
\begin{figure*}[t!]
	\noindent \begin{centering}
		\includegraphics[width=0.95\textwidth]{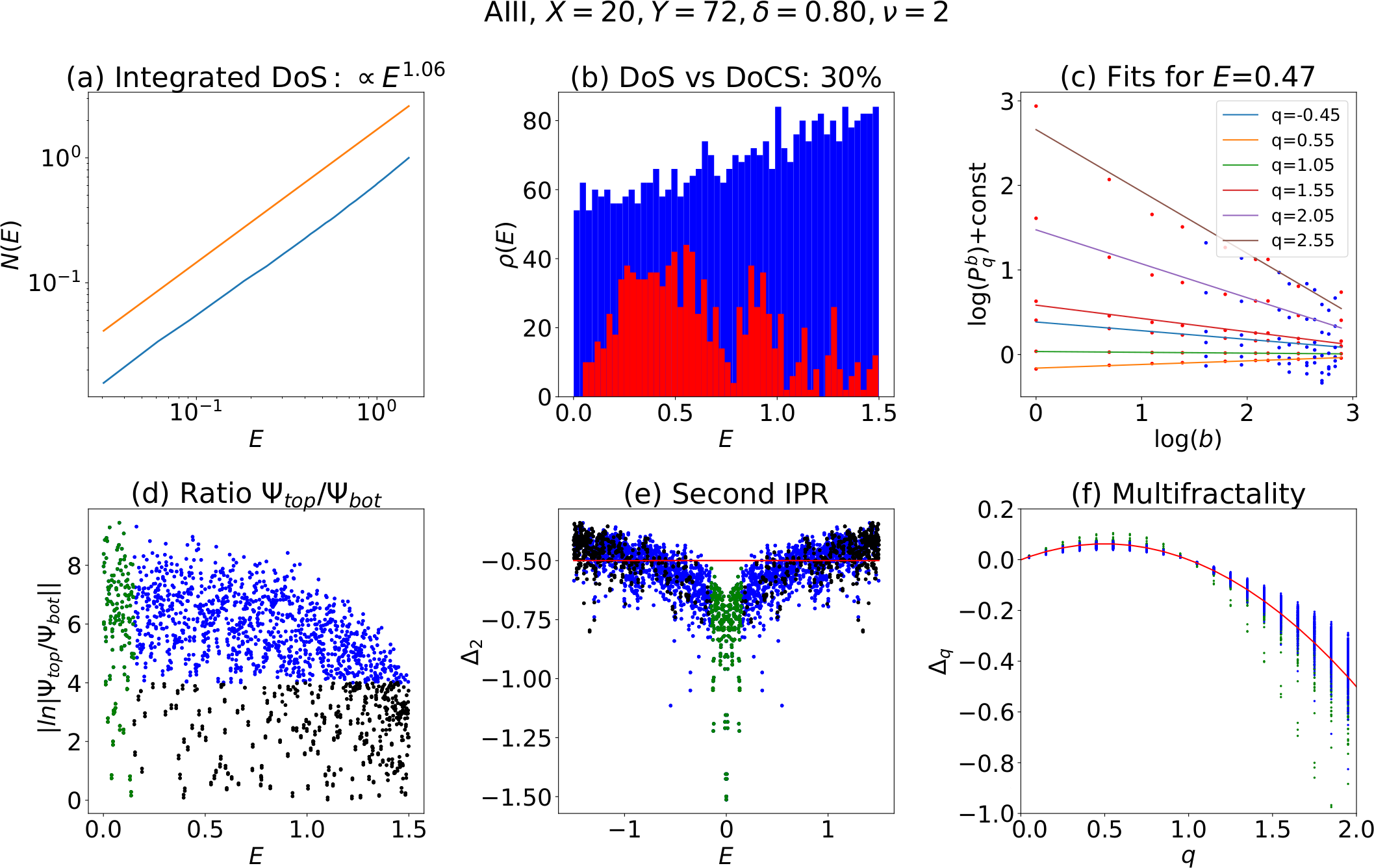}
		\par\end{centering}
	\caption{Numerical exact diagonalization results for the disordered surface states of the 3D class AIII topological superconductor lattice model defined by 
		Eqs.~(\ref{H1})--(\ref{Deltas}). The bulk is tuned to the $\nu = 2$ phase as indicated by the blue dot in Fig.~\ref{fig:lattice_phases}. 
		This figure shows results for a system of surface size $L=72$ with box disorder on all couplings $\lambda \in \{t',\mu_s,t,\Delta\}$ 
		only near the top and bottom slab surfaces.
		For each parameter the disorder spans 
		$\lambda \in [\bar{\lambda} - \delta \, \bar{\lambda},\bar{\lambda} + \delta \, \bar{\lambda}]$, where 
		$\delta = 0.8$
		and 
		the average coupling strengths are 
		$\bar{t'} = 4$, $\bar{\mu}_s = 0$, $\bar{t} = 2$, and $\bar{\Delta} = 1$.	
		Results here characterize eigenstates across the energy ($E$) spectrum for a single, typical realization of the disorder. 
		The energy $E$ is measured in units of $t'$, and the bulk gap is $E = 1.5$.
		(a) The integrated density of states $N(E)$ is shown. Throughout the gap, a power law $N(E) \propto E^\alpha$ with $\alpha = 1.06$ is obeyed. The perfect power law is shown slightly shifted in orange.
		(b) Histogram of the density of critical versus all states. A state is critical, once its 
		anomalous multifractal exponents $\Delta_q$ match the class A QHPT prediction $\Delta_q^\pup{\mathsf{QHPT}} \simeq (0.25) q(1-q)$ \cite{Huckestein1995,Evers2008}
		within 4\% for 75\% of the $q\in[0,q_c]$ below freezing $q_c = 2.83$ \cite{Ghorashi2018,Ghorashi2019}.  
		The total DoS is shown in blue, the critical fraction is colored red. 
		(c) Fits for extracting the multifractal exponents are shown for an exemplary state at $E = 0.47$. Commensurate box sizes are shown as red dots, incommensurate ones (padded with zeros) are blue. 
		When fitting only commensurate boxings are taken into account, and these satisfy the expected linear relation very well even for large $q$. 
		One can observe that $\log P^b_q$ depends linearly on $\log b$ for a single state.
		(d) Vertical depth profile for the surface states. The ratio of the top and bottom surface probabilities 
		[Eq.~(\ref{psiTopBotDef})]
		is plotted versus energy.
		The vast majority are localized on either one of the two faces with open boundary conditions, hence the width of the sample is sufficient to study surface properties.
		States that extend into the bulk are colored black, high-energy surface states blue and low-energy surface states green. 
		This color code is also used in (e),(f). At $E=1.5$, where the gap closes almost all states extend into the bulk.
		(e) Second anomalous dimension $\Delta_2$ as function of energy. For the high-energy 
		surface
		states ($E>0.3$, blue) there is reasonable agreement with $\Delta_2^\pup{\mathsf{QHPT}} = -0.5$.
		(f) Multifractal spectra for all surface states. 
		There is good agreement of the high energy states (blue dots) with the expected 
		QHPT parabola \cite{Huckestein1995,Evers2008} (red line). 
		The low energy states are strongly multifractal (green).}
	\label{fig:lattice_result}
\end{figure*}

\begin{figure}
	\centering
	\includegraphics[width=\columnwidth]{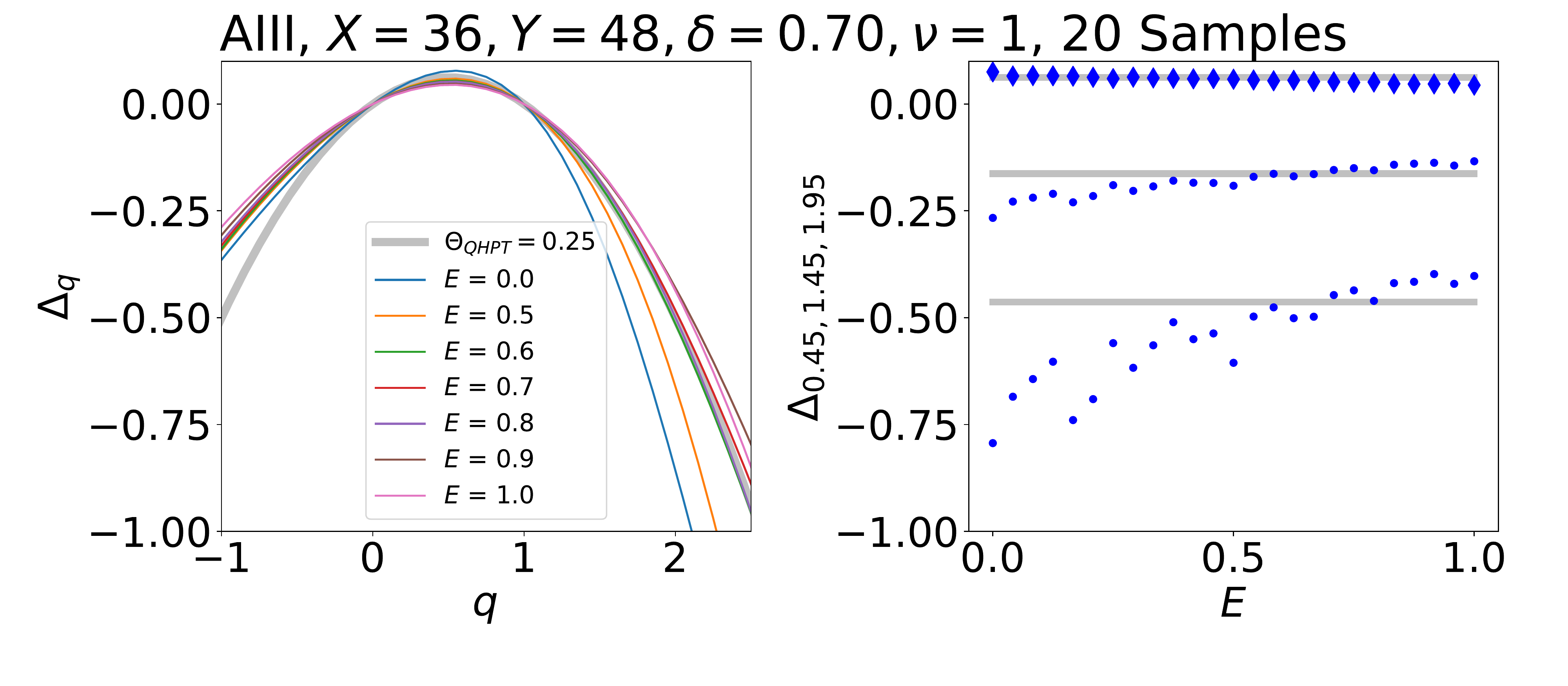}
	\includegraphics[width=\columnwidth]{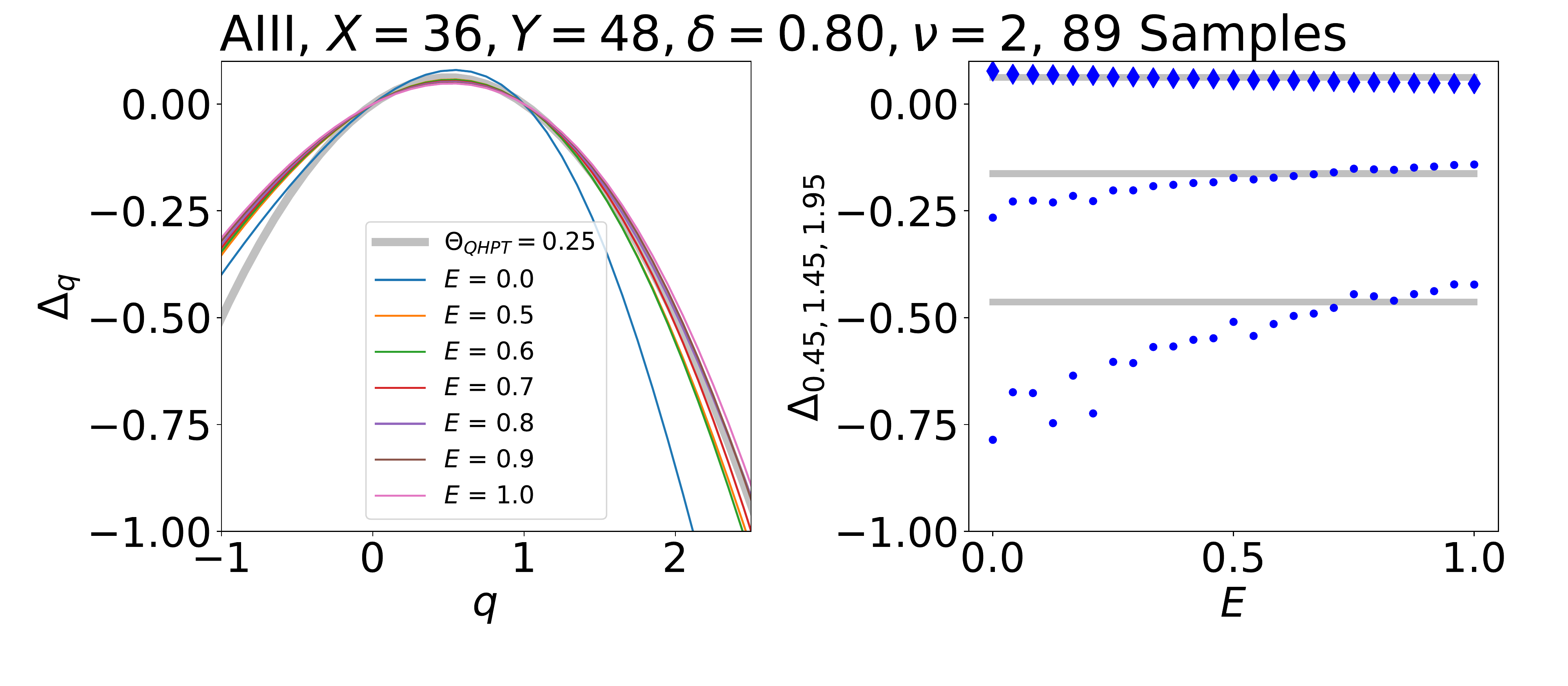}
	\caption{Disorder-averaged multifractal spectra for the lattice model defined by 
	Eqs.~(\ref{H1})--(\ref{Deltas}) with
	surface size $L=48$, slab depth $W=36$, and $\nu=1, 2$. 
	The corresponding bulk topological phases are indicated by the blue dots in Fig.~\ref{fig:lattice_phases},
	while $\bar{t'} = 4$  and $\bar{\Delta} = 1$ as in Fig.~\ref{fig:lattice_result}. 
	For selected energies, we show $\Delta_q$ as a function of $q$ in the left-hand panels. 
	The QHPT parabola $\Delta_q^\pup{\mathsf{QHPT}} \simeq (0.25) q (1-q)$ is shown (thick grey) for comparison. 
	For $\nu=1,2$, selected dimensions 
	$\Delta_{\{0.45,1.45,1.95\}}$ are plotted as a function of energy $E$ on the right. 
	As a guide to the eye, 
	$\Delta^\pup{\mathsf{QHPT}}_{\{0.45,1.45,1.95\}} = \{0.06,-0.16,-0.46\}$ 
	are marked by thick grey lines.
	At zero energy, there is strong multifractality. The gap closes at 
	$E \simeq 1.5$ (not shown) in both cases. 
	The key takeaway is that we observe no even-odd effect: both $\nu = 1,2$ exhibit critical surface
	states (with approximate QHPT multifractality) spanning the bulk energy gap.
	An even-odd effect would mean critical (Anderson localized) finite-energy states for odd (even)
	winding numbers. Such an effect is predicted from a naive deformation of the WZNW term in the theory governing 
	zero-energy critical states, see Sec.~\ref{sec:sigma}. 
	The ED studies on the lattice model agree with $S$-matrix and momentum space (Secs.~\ref{sec:nu=1}, \ref{sec:nu=2}) studies 
	in that both $\nu=1,2$ match the QHPT critical behavior at finite energies.}
	\label{fig:lattice_spectra}
\end{figure}

\section{Dirty surface states of a class AIII 3D topological lattice model \label{sec:lattice}}

We complement the results from the former sections dealing with 
pure
surface theories by exact-diagonalization studies of a slab topological superconductor system. 
We investigate the multifractal spectra of surface states in the superconducting gap.
For this purpose, we take the model from Ref.~\cite{Xie2015}. 
Although not microscopically realistic, this model encodes 
a class AIII topological superconductor on the
diamond lattice, 
as explicated in Fig.~\ref{fig:diamond-lattice} and defined below.
By tuning of parameters one can realize winding numbers of $\nu=-2,-1,0,1,2$. 
See Fig.~\ref{fig:lattice_phases} for a sketch of the topological phase diagram. 
This means we can study both even and odd winding numbers by tuning parameters in this model. 

The lattice Hamiltonian can be decomposed into 3 parts,
$
	H 
	=
	H_{1} + H_{2} + H_{3}.
$
We transcribe the model in the absence of disorder as follows (see also Fig.~\ref{fig:diamond-lattice}).
The first kinetic term consists of nearest-neighbor hopping from one sublattice ($A$) to the other ($B$):
\begin{align}\label{H1}
	\!\!\!
	H_{1} 
	=&\,
	\sum_{\vex{R},\vnn}
	\sum_{\nu=\uparrow,\downarrow}
	t'
	\left[
		C^\dagger_{A\nu}(\vex{R}) 
		\, 
		C_{B\nu}(\vex{R} + \vnn) 
		+ 
		\textrm{H.c.} 
	\right]\!,
	\!\!
\end{align}
where $C_{i \nu}(\vex{R})$ annihilates an electron at site $\vex{R}$
on sublattice $i \in \{A,B\}$, with spin polarization $\nu \in \{\uparrow,\downarrow\}$.  
Here the displacement vectors
$\{\vnn\}$ connect nearest-neighbor $A$ and $B$ sites,
and ``H.c.'' is the Hermitian conjugate. 

A second kinetic term involves 
a staggered chemical potential and next-nearest-neighbor hopping in the 
$xz$- and $yz$-planes,  
\begin{align}\label{H2}
	H_{2} 
	=&\, 
	\!\!
	\sum_{i = A,B}
	\sum_{\vex{R}}
	\sum_{\nu=\uparrow,\downarrow}
	V_{i}
	\,
	C^\dagger_{i\nu}(\vex{R})
	\,
	C_{i\nu}(\vex{R})
\nonumber\\
	&\,
	\!\!
	+
	\sum_{i = A,B}
	\sum_{\vex{R},\vnnn}
	\sum_{\nu=\uparrow,\downarrow}
	\!
	t_i(\vnnn)	
	\!
	\left[
	\begin{aligned}
		&\,
		C^\dagger_{i\nu}(\vex{R})
		\,
		C_{i\nu}(\vex{R} + \vnnn) 
		\\&\,
		+
		\textrm{H.c.}
	\end{aligned}
	\right]\!\!,\!\!\!
\end{align}
where the vectors $\{\vnnn\}$ connect next-nearest-neighbor 
(same-sublattice) sites on the diamond lattice, 
and where 
\begin{align}\label{ViDef}
	V_{i} 
	=&\, 
	\begin{cases}
	\mu_s 	&	i \in A,
	\\
	-\mu_s 	& 	i \in B,
	\end{cases}
\end{align}
\begin{align}\label{tiDef}
	t_i(\vnnn)
	&= \begin{cases}
	t/2, 	& (i \in A, \vnnn \perp \hat{y}) \vee (i \in B, \vnnn \perp \hat{x}),\\
	-t/2, 	& (i \in A, \vnnn \perp \hat{x}) \vee (i \in B, \vnnn \perp \hat{y}),\\
	0, 		& \vnnn \perp \hat{z}.
\end{cases}
\end{align}
The Hamiltonian in Eqs.~(\ref{H1}) and (\ref{H2}) preserves both time-reversal and spin SU(2) symmetries. 
Superconductivity is introduced via 
\begin{align}
\label{H3}
	\!\!\!
	H_{3} 
	=&\, 
	\!\! 
	\sum_{i = A,B}
	\sum_{\vex{R},\vnnn}
	\sum_{\stackrel{\scriptstyle{\mu,\nu}}{=\uparrow,\downarrow}}
	\left[
	\begin{aligned}
	&	\frac{1}{2}
		\Delta_{\mu\nu}(\vnnn)
		\,
		C_{i\mu}^\dagger(\vex{R})
		\,
		C_{i\nu}^\dagger(\vex{R} + \vnnn)
	\\&\,
		+
		\textrm{H.c.}
	\end{aligned}
	\right]
	\!,\!\!
\end{align}
incorporating spin-singlet $d$-wave and spin-triplet $p$-wave pairing terms, 
\begin{align}\label{Deltas}
	\hat{\Delta}(\vnnn)
	=&\, 
	\begin{cases}
	i\sqrt{3} \Delta \hat{\mu}_2 				& (\text{$d$-wave spin singlet}), \;\, \vnnn \perp \hat{z}, 		\\			
	-i \, \sgn(\tilde{\delta}_y) \Delta \hat{\mu}_1 	& (\text{$p$-wave spin triplet}), \;\, \vnnn \perp \hat{x},		\\
	-i \, \sgn(\tilde{\delta}_x) \Delta \hat{\mu}_1 	& (\text{$p$-wave spin triplet}), \;\, \vnnn \perp \hat{y}.
\end{cases}
\end{align}
In the last equation, the Pauli matrices $\{\hat{\mu}_{1,2,3}\}$ act on spin-1/2 space. 

For nonzero $\{t',\Delta\}$, 
the full BdG Hamiltonian
in Eqs.~(\ref{H1})--(\ref{Deltas}) is fully gapped in the bulk, except for $\{\mu_s,t\}$ constrained 
along the topological phase
transition lines indicated in Fig.~\ref{fig:lattice_phases}. 
It preserves time-reversal $\mathcal{T}$ symmetry, as well as invariance under U(1) rotations about the $\hat{z}$-axis in spin ($\{\uparrow,\downarrow\}$) space \cite{Xie2015}.
With superconductivity, these symmetries realize class AIII \cite{Foster2008,SRFL2008}. 

We implement the Hamiltonian for a slab geometry. 
In the $x$-direction 
we terminate the diamond lattice in two surfaces,
with $W$ pairs of $(L\times L)$ $A,B$ layers in between. 
We choose periodic boundary conditions in $y$- and $z$-directions to obtain a quasi-infinite slab of thickness $W$. 
Disorder is implemented by choosing each coupling $\lambda \in \{t',\mu_s,t,\Delta\}$ [Eqs.~(\ref{H1})--(\ref{Deltas})] 
from a box distribution 
$\lambda \in [\bar{\lambda} - \delta_x \, \bar{\lambda},\bar{\lambda} + \delta_x \, \bar{\lambda}]$, where $\bar{\lambda}$ is the 
average value of the coupling strength. 
In order to avoid localization of the bulk states, we choose $\delta_x$ that quickly decays away from the two surfaces, 
so that the disorder effectively resides only near sample boundaries.
Suppressing bulk disorder further prevents shifting of topological phase boundaries as function of the disorder strength.

To obtain all the in gap states of the sparse Hamiltonian, we employ the FEAST algorithm \cite{FeastAlgo}. In numbers, this enables us to compute $\mathcal{O}(10^3)$ eigenstates in the gap for matrix sizes up to $0.5\times 10^6$ to accuracy of $10^{-10}$ times the matrix norm. As one can see by comparing to the data in Refs.~\onlinecite{Evers2001, Obuse2010, Obuse2012} on QH critical wave functions on the network model, we can access the IPRs with reasonable precision using the system sizes under consideration. To investigate parabolicity and critical exponents, i.e.\ accessing more than a decade of systems sizes with averaging, our computational power is insufficient. We therefore leave precision finite size scaling studies of the topological surface wave functions to future work.

The lattice wave function carries 
crystal coordinates $\{x,y,z\}$, 
sublattice index $i \in \{A,B\}$, 
and spin index $\nu \in \{\uparrow,\downarrow\}$. 
The surface-resolved probability density of an eigenstate wave function 
$\psi_{x,y,z;i,\nu}$ is defined via
\begin{align}
	|\psi^S_{y,z}|^2 
	\equiv&\,
	\sum_{i = A,B}
	\sum_{x}
	\sum_{\nu = \uparrow,\downarrow}
	|\psi_{x,y,z;i,\nu}|^2,
\end{align}
where we trace over the slab depth $x$, as well as sublattice and spin spaces.
The surface states decay quickly (exponentially) into the bulk, 
and our results are not dependent on how many slices orthogonal to the $x$-direction are taken 
for states deeply in the gap. 

We define the box probability $\mu_n$ 
and the inverse participation ratio $P_q^b$ as usual,
\begin{align}
	\mu_n 
	=&\, 
	\sum_{y,z \in A^b_n} 
	|\psi^S_{y,z}|^2,
\\
	P_q^{b} 
	=&\, 
	\sum_n \mu_n^q.
\end{align}
We subdivide the $(L \times L)$ 
surface into square boxes $\{A^b_n\}$ 
of linear size $b$. For the linear regression to extract the multifractal exponents 
$\tau_q$, only commensurate box-sizes are taken into account:
\begin{align}
	\ln P_q^{b} 
	= 
	\tau_q\ln b 
	+
	c. 
	\label{eq:tauf}
\end{align}
For $0<q<1$, the restriction to commensurate box-sizes is crucial.

For a large system with 
linear surface size $L = 72$ and
even winding number $\nu=2$, we show 
the density of states and eigenstate multifractality 
in Fig.~\ref{fig:lattice_result}. 
Energy $E$ is measured in units of the nearest-neighbor hopping strength $t'$.
For the analysis presented in Fig.~\ref{fig:lattice_result},
we only extract the in-gap states that comprise less than $1\%$ of the entire 
energy spectrum. Throughout the gap, a power-law 
integrated density of states
$N(E) \propto E^\alpha$, with $\alpha$ close to unity is obeyed [Fig.~\ref{fig:lattice_result}(a)]. 
The dimensionless disorder strength $\delta$ 
implemented on all surface couplings of the microscopic lattice model
is related to the abelian ($\lambda_A$) and nonabelian disorder strengths 
of the surface theory 
[Eqs.~(\ref{hAIII2}) and (\ref{lamADef})]
in a non-trivial way. For this reason,
a comparison of bare and effective disorder via Eq.~(\ref{DoS_nu}) as in the preceding sections 
is therefore not revealing in this case. 

In Fig.~\ref{fig:lattice_result}(b), 
a histogram exhibiting the density of critical versus the density of all states is shown. 
A state is critical, once its anomalous multifractal exponents $\Delta_q$ match the class A QHPT prediction 
$\Delta_q^\pup{\mathsf{QHPT}} \simeq (0.25) q(1-q)$ \cite{Huckestein1995,Evers2008}
within 4\% for 75\% of the $q\in[0,q_c]$ below freezing $q_c = 2.83$ \cite{Ghorashi2018,Ghorashi2019}. 
Almost a third of the in-gap states satisfies this.
We verify that the inverse participation ratios $\log P^b_q$ depend linear on $\log b$ and show this for an exemplary state 
in Fig.~\ref{fig:lattice_result}(c). 
Further in Fig.~\ref{fig:lattice_result}(d), we check that the wave functions in the gap 
consist primarily of surface states
by analyzing their 
behavior integrated over slices in the open direction $x$.
Here, the total surface probabilities are defined via 
\begin{align}\label{psiTopBotDef}
\begin{aligned}
	|\psi_{top}|^2 
	=&\,
	\sum_{i = A,B}
	\sum_{y,z}
	\sum_{\nu = \uparrow,\downarrow}
	|\psi_{x = W,y,z;i,\nu}|^2,
\\
	|\psi_{bot}|^2 
	=&\,
	\sum_{i = A,B}
	\sum_{y,z}
	\sum_{\nu = \uparrow,\downarrow}
	|\psi_{x = 0,y,z;i,\nu}|^2.	
\end{aligned}
\end{align}	
Finally, panels (e) and (f) in Fig.~\ref{fig:lattice_result} exhibit anomalous
multifractal spectra for states spanning the bulk gap. 
Although states near zero energy exhibit stronger multifractality,
most of the in-gap states show weaker fractality,
consistent with the QHPT prediction. 

In Fig.~\ref{fig:lattice_spectra}, we show 
disorder-averaged
multifractal spectra for smaller ($L=48$) systems. We compare the two winding numbers $\nu=1$ and $\nu=2$. 
In agreement with the preceding conductivity and multifractality study of the surface theory (Secs.~\ref{sec:nu=1} and \ref{sec:nu=2}), 
we find the same behavior at finite energies for both winding numbers. In accordance with the expectation 
[Eq.~\eqref{theta_nu}] for the topological AIII surface theory, there is strong multifractality for strong disorder at zero energy. 
For $E > 0.5$ , there is very good agreement between the class A QHPT parabola and the observed average multifractal spectra. 
At higher energies (not shown), the surface states start extending into the clean bulk.


\section{Discussion \label{sec:disc}}

\subsection{Sigma models, absent even-odd effect, and phase diagram \label{sec:sigma}}

The statistics of spatial fluctuations in the 
zero-energy eigenstates of the class AIII topological surface-state Hamiltonians in Eqs.~(\ref{hAIII2}) and 
(\ref{hAIII1}) can be encoded in a matrix sigma model \cite{Evers2008}. The form of the theory
obtains from non-abelian bosonization 
\cite{Ludwig1994,Nersesyan1994,Mudry1996,Caux1996,Guruswamy2000,Bhaseen2001,AltlandSimonsZirnbauer2002} 
and conformal embedding theory \cite{Foster2014},
\begin{align}
	\!\!\!
	S
	=&\,
	\frac{\nu}{8 \pi}
	\int
	d^2\vex{r}
	\,
	\Tr\!\left[
	\Nabla \hat{q}^\dagger 
	\cdot
	\Nabla \hat{q} 
	\right]
\nonumber\\&\,
	-
	\frac{\lambda_A \, \nu^2}{8 \pi^2}
	\int
	d^2\vex{r}
	\,
	\Tr\!\left[
	\hat{q}^\dagger
	\Nabla 
	\hat{q}
	\right]
	\cdot
	\Tr\!\left[
	\hat{q}^\dagger
	\Nabla 
	\hat{q}
	\right]
\nonumber\\&\,
	+
	\frac{i \omega}{2}
	\int
	d^2\vex{r}
	\,
	\Tr\!\left[
	\hat{\Lambda}
	\left(
	\hat{q} + \hat{q}^\dagger
	\right)
	\right]
\nonumber\\&\,
	-
	\frac{i \nu}{12 \pi}
	\int
	d^2\vex{r}
	\,
	d R
	\,
	\epsilon^{a b c}
	\Tr\!\left[
	(\hat{q}^\dagger \partial_a \hat{q})
	(\hat{q}^\dagger \partial_b \hat{q})
	(\hat{q}^\dagger \partial_c \hat{q})
	\right]\!.
	\!\!\!
\label{WZNW}
\end{align}
Here $\nu$ is the bulk winding number, 
equal to one (two) for the Hamiltonian in 
Eq.~(\ref{hAIII1}) [(\ref{hAIII2})]. 
The parameter $\lambda_A$ is the strength
of the abelian disorder potential 
[Eq.~(\ref{lamADef})].
The parameter $\omega$ is the ac frequency 
at which the conductivity is to be evaluated.
Taking $\omega \neq 0$ probes states away from zero energy.  

The field $\hat{q}(\vex{r})$ is a $(2 n)$$\times$$(2 n)$ unitary matrix
[$\hat{q}^\dagger(\vex{r}) \, \hat{q}(\vex{r}) = \hat{1}_{2n}$, where $\hat{1}_{2n}$ denotes the $(2 n)$$\times$$(2 n)$ identity],
with $n \rightarrow 0$ equal to the number of fermionic replicas \cite{Evers2008,Foster2014}. 
The frequency $\omega$ couples to the (\emph{imaginary} \cite{AltlandSimonsZirnbauer2002}) ``mass''
term $(i/2) \Tr[\hat{\Lambda}(\hat{q} + \hat{q}^\dagger)]$, where $\hat{\Lambda} = \diag{\{\hat{1}_n,-\hat{1}_n\}}$ 
grades in retarded/advanced space \cite{Evers2008}. 
The fourth line in Eq.~(\ref{WZNW}) is the Wess-Zumino-Novikov-Witten term,
which requires the extension of the field domain to the 3-ball $(R,\vex{r})$, with $0 \leq R \leq 1$. 
The third coordinate $R$ can be thought of as accessing the ``bulk'' of the topological phase,
with $R = 1$ denoting the surface \cite{Koenig2012}. 

The first three lines of Eq.~(\ref{WZNW}) describe the topologically trivial 
``Gade'' AIII Dirac model \cite{Guruswamy2000,DellAnna2006,Foster2008,Koenig2012}, 
defined by the restriction 
of Eq.~(\ref{hDef}) via the sublattice (chiral) symmetry in Eq.~(\ref{SDef}). 
As we have seen in Sec.~\ref{sec:nontop} and Fig.~\ref{fig:nu2-localizing}, 
the finite-energy states of this model are strongly Anderson localized. 

It is the advent of the WZNW term on the fourth line of Eq.~(\ref{WZNW}) 
that distinguishes the topological class AIII models 
[Eqs.~(\ref{hAIII2}) and (\ref{hAIII1})], which are protected by the \emph{anomalous} chiral symmetry in Eq.~(\ref{STopDef}).
The key zero and low-energy properties of topological class AIII surface states, 
summarized in Eqs.~(\ref{theta_nu}), (\ref{z}), (\ref{DoS_nu}), and (\ref{CondWZNW}), can be derived from 
Eq.~(\ref{WZNW}) with $\omega = 0$ \cite{Ludwig1994,Nersesyan1994,Mudry1996,Caux1996,Guruswamy2000,Bhaseen2001,Tsvelik1995,Ostrovsky2006,Foster2014}. 
We have numerically established that these models exhibit ``stacks'' of critically delocalized QHPT states at finite energy in 
Secs.~\ref{sec:nu=1}, \ref{sec:nu=2}, and \ref{sec:lattice}.

\begin{figure}
	\includegraphics[width=0.95\columnwidth]{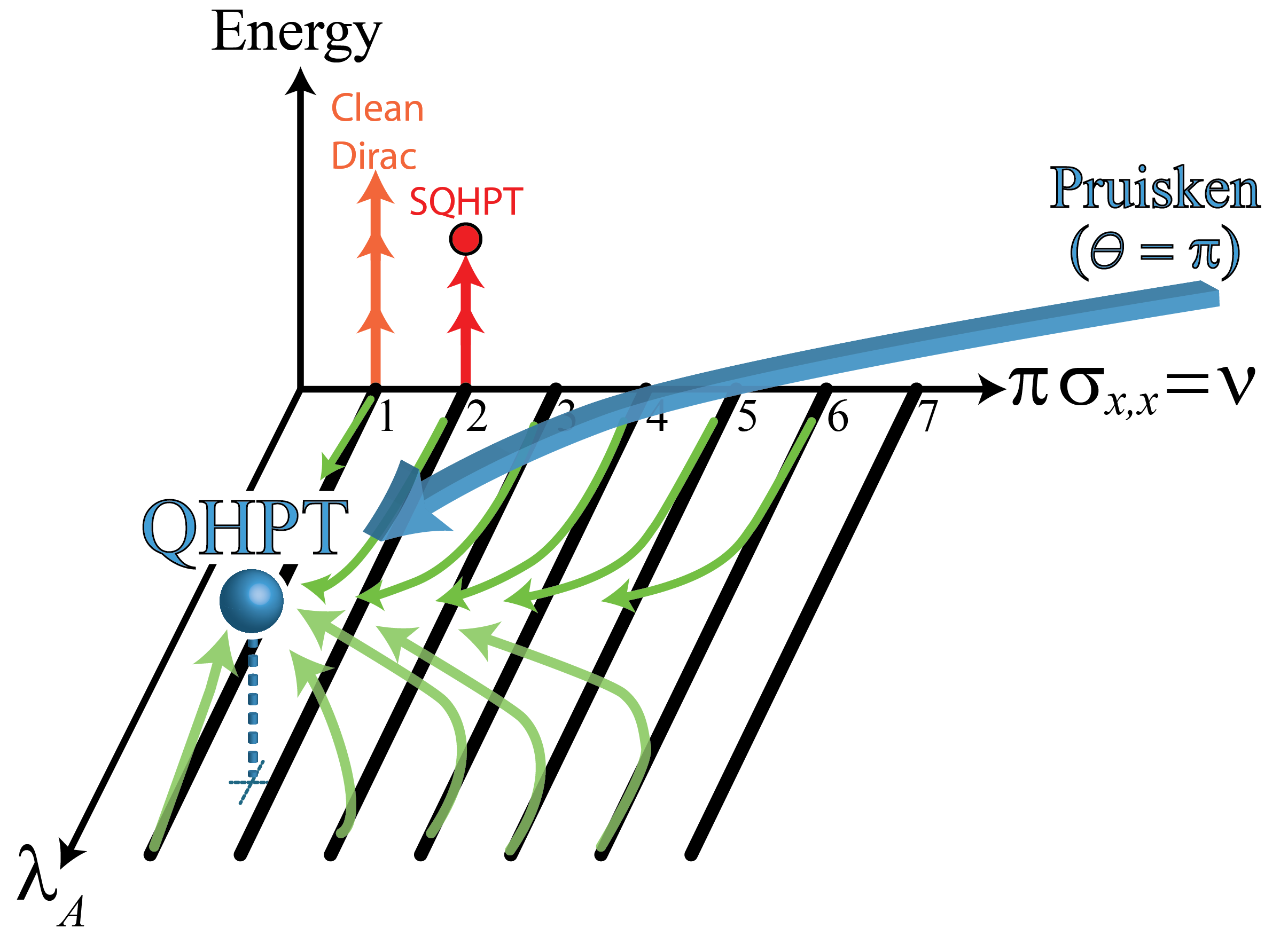}
	\caption{Schematic ``physical'' renormalization group flow 
	picture connecting zero-energy and finite-energy fixed-point descriptions
	of class AIII topological surface states. 
	In the zero-energy horizontal plane, the system state is determined
	by the winding number $\nu$ [equal to $\pi$ times the longitudinal conductivity,
	Eq.~(\ref{CondWZNW})] and by the abelian disorder strength $\lambda_A$, 
	Eqs.~(\ref{lamADef}) and (\ref{WZNW}). The latter parameterizes 
	the parallel zero-energy fixed lines, indicated in black. 
	The numerical results presented in Secs.~\ref{sec:nu=1}, \ref{sec:nu=2}, and
	\ref{sec:lattice} strongly suggest that nonzero energy induces a flow
	(green arrows) to the logarithmic conformal field theory that governs the plateau
	transition of the ordinary quantum Hall effect (QHPT). 
	For $\lambda_A = 0$, finite energy produces special behavior for 
	winding numbers $\nu = 1$ (clean Dirac) and $\nu = 2$. The latter
	flows instead to the plateau transition of the \emph{spin} quantum Hall plateau
	transition (SQHPT) \cite{Ghorashi2018}. 
	Also indicated is the schematic flow to the QHPT critical point 
	from the ultraviolet limit of the quantum Hall plateau transition, 
	as captured by the Pruisken sigma model in Eq.~(\ref{Pruisken}) with
	a large bare conductivity $\sigma_{x,x} \gg 1$ and half-integer $\sigma_{x,y}$. 
	}
	\label{fig:RGPD}
\end{figure}

A symmetry-based argument for the effect of $\omega \neq 0$ in Eq.~(\ref{WZNW}) is the following.
With $\omega = 0$, the model exhibits U$(2n)$$\otimes$U$(2n)$ symmetry,
defined as invariance under independent left- and right- group transformations,
\begin{align}
	\hat{q} \rightarrow \hat{U}_L \, \hat{q} \, \hat{U}_R^\dagger,
	\quad
	\hat{U}_L^\dagger \hat{U}_L
	=
	\hat{U}_R^\dagger \hat{U}_R
	=
	\hat{1}_{2n}. 
\end{align}
After absorbing the matrix $\hat{\Lambda}$ by left-group translation, 
$\omega \neq 0$ in Eq.~(\ref{WZNW}) constrains $\hat{U}_L = \hat{U}_R$, 
reducing the symmetry to the diagonal U($2n$). 
Although $\omega$ constitutes an imaginary mass, if we assume
that the oscillations induced by fluctuations in the larger U$(2n)$$\otimes$U$(2n)$
space are suppressed \cite{AltlandSimonsZirnbauer2002}, we can impose ``by hand'' the constraint
\begin{align}\label{ClassAConstraint}
	\hat{q} = \hat{q}^\dagger,
	\quad
	\Tr\left[\hat{q}\right] = 0.
\end{align}
Then $\hat{q}^2 = \hat{1}_{2n}$.
Imposing these conditions in the presence of the WZNW term requires care, because
the constraint introduces topologically distinct sectors of $\hat{q}(\vex{r})$. 
Using a standard deformation of the WZNW term due to Bocquet, Serban, and Zirnbauer 
\cite{D-2_BocquetZirnbauer2000,AltlandSimonsZirnbauer2002,Konig2014}, the sigma 
model in Eq.~(\ref{WZNW}) reduces to a version of Pruisken's model
for the quantum Hall effect \cite{Pruisken1987}, 
\begin{align}
	S
	\rightarrow
	&\,
	\frac{\sigma_{x,x}}{8}
	\int
	d^2\vex{r}
	\,
	\Tr\!\left[
	\Nabla \hat{q}
	\cdot
	\Nabla \hat{q} 
	\right]
\nonumber\\&\,
	-
	\frac{\sigma_{x,y}}{8}
	\int
	d^2\vex{r}
	\,
	\epsilon^{i j}
	\Tr\!\left[
	\hat{q} 
	\, 
	\partial_i \hat{q}
	\,
	\partial_j \hat{q}
	\right]\!,
\label{Pruisken}
\end{align}
where 
\begin{align}\label{PruiskenParams}
	\sigma_{x,x} = \nu/\pi,
	\quad
	\sigma_{x,y} = \nu/2.
\end{align}
The $\lambda_A$ and frequency terms have now disappeared due to Eq.~(\ref{ClassAConstraint}).
[We perform the group translation that replaces $\hat{\Lambda} \rightarrow \hat{1}_{2n}$ in Eq.~(\ref{WZNW}),
before implementing the constraint.] 

The topological theta term on the second line of Eq.~(\ref{Pruisken})
has the coefficient $\sigma_{x,y}$, given by Eq.~(\ref{PruiskenParams}).
In the context of a single massless Dirac fermion, it has 
been argued that the parameter $\sigma_{x,y}$ is \emph{not} the physical
Hall conductivity \cite{Konig2014}.
The latter should vanish for the on-average parity-invariant
class AIII surface \cite{AltlandSimonsZirnbauer2002}.

The main point is that Eq.~(\ref{PruiskenParams}) implies
a nontrivial even-odd effect. For odd values of the winding number
$\nu$, the Pruisken term in Eq.~(\ref{Pruisken}) has $\vartheta \equiv 2 \pi \sigma_{x,y} = \pi$ (modulo $2 \pi$),
corresponding to the plateau transition of the integer quantum Hall effect \cite{Ostrovsky2007,Evers2008,Pruisken1987}. 
On the other hand, the parameter $\vartheta = 2 \pi$ for even winding numbers.
This corresponds to the Anderson insulating quantum Hall plateau. 
We conclude that the ``derivation'' of Eq.~(\ref{Pruisken}) from 
Eq.~(\ref{WZNW}) predicts that class AIII surface states at finite energy 
are critically delocalized at the QHPT (Anderson localized in the plateau) for odd (even) 
winding numbers. Although we have obtained Eqs.~(\ref{Pruisken}) and (\ref{PruiskenParams}) using
nonabelian bosonization, the same conclusion is seemingly implied by a
straight-forward gradient expansion \cite{Ostrovsky2007,Essin2015}. 

A key result of the numerical results presented in Secs.~\ref{sec:nu=1}, \ref{sec:nu=2}, 
and \ref{sec:lattice} is that no such even-odd effect is observed. 
The Landauer conductivity and multifractal spectra results for surface theories with $\nu = 1,2$ 
shown in Figs.~\ref{fig:nu1} and \ref{fig:nu2}, as well as the results for the surface states
of the class AIII bulk lattice model exhibited in Fig.~\ref{fig:lattice_spectra}, show no difference
between even and odd winding numbers. In both cases, finite-energy states
are observed to form a ``stack'' of quantum critical QHPT states. 
 
We emphasize that Eqs.~(\ref{Pruisken}) and (\ref{PruiskenParams}) 
do not obtain by following a physical renormalization group (RG) flow,
starting from the $\omega \neq 0$ perturbed WZNW model in Eq.~(\ref{WZNW}). 
Instead, the imposition of the constraint in Eq.~(\ref{ClassAConstraint})
is essentially a ``mean field'' guess for how the zero-energy field
theory responds to the finite-energy perturbation. 
It is certainly possible that a physical RG flow connects Eq.~(\ref{WZNW}) 
with $\omega \neq 0$ in the ultraviolet to Eq.~(\ref{Pruisken}),
but with $\sigma_{x,y} = 1/2$ for all nonzero $\nu$. 
At the same time, the Pruisken theory is also an ultraviolet
description of quantum Hall physics, i.e.\ Eq.~(\ref{Pruisken}) can only
be derived in a controlled fashion for large bare $\sigma_{x,x} \gg 1$ \cite{Pruisken1987,Evers2008}. 
The Pruisken model flows to strong coupling $\sigma_{x,x} \rightarrow \ord{1}$;
for $\sigma_{x,y} = 1/2$, this flow is presumed to terminate
at the logarithmic conformal fixed point governing the QHPT.

A schematic RG flow diagram is shown in Fig.~\ref{fig:RGPD}, connecting 
the zero-energy fixed lines of Eq.~(\ref{WZNW}) (with $\omega =0$, 
parameterized by the purely marginal abelian disorder variance $\lambda_A$),
to the putative QHPT critical point. A conjectured flow from the ultraviolet 
Pruisken model (with $\vartheta = \pi$) is also indicated in this figure.

\subsection{Critical stacking and Zirnbauer's $\nu = 4$ AIII model for the QHPT \label{sec:Zirnbauer}}

Both the class AIII WZNW model in Eq.~(\ref{WZNW}) and the class A Pruisken model in Eq.~(\ref{Pruisken}) 
model are defined by U$(2n)$ unitary symmetry. In fact, for the special case of one replica $n = 1$, 
these theories can become equivalent. The Pruisken model with $n = 1$ has target manifold
\[
	\frac{\text{U}(2)}{\text{U}(1)\otimes\text{U}(1)}
	\simeq
	\frac{\text{O}(3)}{\text{O}(2)}.
\]
In this case, Eq.~(\ref{Pruisken}) is the effective Euclidean spacetime field theory for the 1+1-D
antiferromagnetic SU(2) Heisenberg spin-$s$ chain, with $\sigma_{x,y} = s$ \cite{FradkinBook}. 
According to Haldane's conjecture, for half-integer $s$ 
this model is equivalent to the SU(2) sector of the U(2) 
WZNW model in Eq.~(\ref{WZNW}), at level $\nu = 1$. 
Although the ``ultraviolet'' 
Pruisken theory has only diagonal SU(2) symmetry, the WZNW model is invariant under the larger chiral symmetry group SU(2)$\otimes$SU(2). 
Enlarged chiral symmetry is a generic property of rational 2D conformal field theories. 
By contrast, spin chains with integer $s$ have $\vartheta = 2\pi$, 
and flow to a gapped paramagnetic phase: this is the famous even/odd effect 
predicted by Haldane's conjecture \cite{FradkinBook}. 

Although the replica limit $n \rightarrow 0$ produces nonunitary, $c = 0$ logarithmic conformal
field theories \cite{Bhaseen2001,Cardy2013}, one might hope that a version of the Haldane conjecture 
survives in the replica limit.
The QHPT features a noncritical density of states, 
as well as an approximately parabolic multifractal spectrum $\tau(q)$, Eq.~(\ref{tau(q)}) with 
$\theta \simeq 0.25$ \cite{Huckestein1995,Evers2008}. 
We can apparently satisfy both of these conditions in the class AIII WZNW
model [Eqs.~(\ref{theta_nu}), (\ref{z}) and (\ref{DoS_nu})] by choosing 
\begin{align}\label{Zirnnu4}
	\frac{\lambda_A}{\pi} = \frac{1}{\nu^2}, \quad \nu = 4.
\end{align}

Very recently, Zirnbauer has argued that consideration of
the Chalker-Coddington network model independently implies
the existence of a class AIII current algebra with level $\nu = 4$ \cite{Zirnbauer2019}.
Owing to the special way that the WZNW model emerges in the quantum Hall context, 
Zirnbauer argues that the mean conductance at the plateau transition
is exactly half the predicted value given by Eq.~(\ref{CondWZNW}),
$\sigma_{x,x} = 2/\pi \simeq 0.64$. 
This is not too different from previous numerics [Eq.~(\ref{CondQHPT})]  \cite{Huckestein1995,Huo1993,Cho1997,Wang1998,Schweitzer2005,Evers2008},
and the value we find at finite energy for $\nu = 1,2$ class AIII
topological surface states in Figs.~\ref{fig:nu1}, \ref{fig:nu2}, and \ref{fig:nu2-WNsweep}. 

A very surprising aspect of Zirnbauer's proposal \cite{Zirnbauer2019} 
is that the correlation length exponent $\nuqh$ 
(not to be confused with the class AIII winding number $\nu$)
for tuning away from the transition into the quantum Hall plateau
is actually predicted to be \emph{infinite}.
The claim is that existing numerical studies showing $\nuqh$ in the range between $2.3$--$2.6$
\cite{Huckestein1995,Evers2008,Slevin2009,Obuse2010,Amado2011,Dahlhaus2011,Fulga2011,Obuse2012,Slevin2012,Nuding2015,Gruzberg2017}
are beset by finite-size effects. 
We emphasize that for the finite-energy class AIII surface states that exhibit QHPT critical statistics studied here,
the exponent $\nuqh$ does not play a role. This is because all states are delocalized,
and the only relevant length scale is the \emph{crossover scale} $\zeta(E) \sim \e^{-1/z}$ between
the low-energy class AIII states (which exhibit winding number $\nu$-dependent statistics) 
and the finite-energy QHPT states. Here $z$ is the dynamic critical exponent 
associated to the class AIII, low-energy states [Eq.~(\ref{z})].

The numerical results obtained in this paper open up a potential avenue to 
test Zirnbauer's $\nu = 4$ theory. In particular, if we assume that the $\nu = 4$ 
class AIII WZNW model [Eq.~(\ref{WZNW})] also exhibits a ``stack'' of QHPT critical
eigenstates at finite energy, then the only difference between zero- and finite-energies
\emph{for $\nu = 4$} is the fine-tuning of the abelian parameter $\lambda_A$.
The latter must satisfy Eq.~(\ref{Zirnnu4}) in order to realize QHPT multifractality at
finite energy. Thus, both zero- and finite-energy states would be governed
by the same quantum field theory, and the only missing piece is a controlled
renormalization group mechanism that fine tunes $\lambda_A \rightarrow \pi/16$ 
for $\omega \neq 0$ in Eq.~(\ref{WZNW}). This is an interesting avenue
for future analytical and/or numerical work.

\subsection{Conclusion}

In this paper we have presented substantial numerical evidence that \emph{most} finite-energy surface
states of a class AIII topological superconductor (or chiral topological insulator) 
form a ``stack'' of quantum critical, QHPT states, in the presence of surface disorder
that preserves the defining class AIII symmetries. 
Similar results based on the multifractal analysis of surface states with disorder were obtained in Refs.~\cite{Ghorashi2018,Ghorashi2019},
for class CI and DIII topological superconductors. 
For class AIII treated here, we have provided two additional sources of evidence for the stacking 
of quantum Hall plateau transition states: 
(1) Landauer conductance of the finite-energy states in the 
continuum surface-only theory, and 
(2) multifractal analysis of all in-gap states of a class AIII
topological diamond lattice model in the slab geometry. 

In total, the results here and in Refs.~\cite{Ghorashi2018,Ghorashi2019} strongly argue for
an unexpected, new connection between $\mathbb{Z}$-graded topological phases in two and three spatial dimensions.
The 2D topological phases in classes C, A, and D describe the spin, charge (ordinary integer), and thermal quantum Hall effects,
with broken time-reversal symmetry. 
The 3D topological phases in classes CI, AIII, and DIII can describe time-reversal invariant topological superconductors.  
The numerics suggest that the quantum critical plateau transitions in classes C, A, and D, which occur only
at isolated energies in two dimensions, reappear in gap-spanning ``stacks'' at the surfaces of class CI, AIII,
and DIII topological phases, respectively, in the presence of symmetry-preserving disorder. 

Much of the physics of these newly uncovered quantum-critical, multifractal surface metals remains to be explored.
The role of interparticle interaction-mediated instabilities in the presence of gap-spanning multifractality 
is an obvious example. Another avenue is the investigation of Zirnbauer's proposal for the field theory of 
the plateau transition as a particular \emph{zero-energy} class AIII topological surface state \cite{Zirnbauer2019},
see Sec.~\ref{sec:Zirnbauer} for a discussion. 

Classes CI, AIII, and DIII are also notable for exhibiting a precisely quantized longitudinal surface spin or thermal
conductivity, which holds even in the presence of disorder and interactions \cite{Ludwig1994,Tsvelik1995,Ostrovsky2006,Evers2008,Xie2015}.
The connection between 2D class C, A, and D quantum Hall effects and 3D class CI, AIII, and DIII topological superconductors 
likely reflects a deep, topological origin. The latter remains to be uncovered in future studies.

\acknowledgements
We thank 
Alexander Altland, 
Ilya Gruzberg, 
Victor Gurarie,
and 
Alexander Mirlin for useful discussions. We thank Ziqiang Wang for letting us use the conductance distribution data of Ref.\ \cite{Jovanovic1998}.
This research was supported by NSF CAREER Grant No.~DMR-1552327 (MSF), the KHYS research travel grant and Graduate Funding from the German States (JFK)
and the German National Academy of Sciences Leopoldina through grant LPDS 2018-12 (BS).


\begin{thebibliography}{99}
\bibitem{KaneHasan}
	M. Z. Hasan and C. L. Kane, 
	\emph{Colloquium: Topological insulators,}
	Rev. Mod. Phys. {\bf 82}, 3045 (2010).
\bibitem{TSCRev1}
	X.-L. Qi and S.-C. Zhang,
	\emph{Topological insulators and superconductors,}
	Rev. Mod. Phys. {\bf 83}, 1057 (2011).
\bibitem{BernevigBook}
	B. A. Bernevig and T. L. Hughes,
	\emph{Topological Insulators and Topological Superconductors}
	(Princeton University Press, Princeton, New Jersey, 2013).
\bibitem{SRFL2008}
	A. P. Schnyder, S. Ryu, A. Furusaki, and A. W. W. Ludwig, 
	\emph{Classification of topological insulators and superconductors in three spatial dimensions,} 
	Phys. Rev. B {\bf 78}, 195125 (2008).
\bibitem{Essin2015} 
	A. M. Essin and V. Gurarie, 
	\emph{Delocalization of boundary states in disordered topological insulators,}
	J. Phys. A: Math. Theor. \textbf{48} (2015).
\bibitem{Essin2011}
	A. M. Essin and V. Gurarie, 
	\emph{Bulk-boundary correspondence of topological insulators from their respective Green's functions,}
	Phys. Rev. B {\bf 84}, 125132 (2011).
\bibitem{Evers2008}
	F. Evers and A. D. Mirlin,
	\emph{Anderson transitions,}
	Rev. Mod. Phys. {\bf 80}, 1355 (2008).
\bibitem{LeeRamakrishnan1985}
	P. A. Lee and T. V. Ramakrishnan,
	\emph{Disordered electronic systems,}
	Rev. Mod. Phys. {\bf 57}, 287 (1985).
\bibitem{KaneMele2005}
	C. L. Kane and E. J. Mele,
	\emph{Quantum Spin Hall Effect in Graphene,}
	Phys. Rev. Lett. {\bf 95}, 226801 (2005).
\bibitem{Xie2016}
	H.-Y.  Xie, H. Li, Y.-Z. Chou, and M. S. Foster,
	\emph{Topological Protection from Random Rashba Spin-Orbit Backscattering: Ballistic Transport in a Helical Luttinger Liquid,}
	Phys. Rev. Lett. {\bf 116}, 086603 (2016).
\bibitem{Bardarson2007}
	J. H. Bardarson, J. Tworzydlo, P. W. Brouwer, and C. W. J. Beenakker,
	\emph{One-Parameter Scaling at the Dirac Point in Graphene,}
	Phys. Rev. Lett. {\bf 99}, 106801 (2007).
\bibitem{Nomura2007}
	K. Nomura, M. Koshino, and S. Ryu,
	\emph{Topological Delocalization of Two-Dimensional Massless Dirac Fermions,}
	Phys. Rev. Lett. {\bf 99}, 146806 (2007).
\bibitem{Ryu2007}
	S. Ryu, C. Mudry, H. Obuse, and A. Furusaki,
	\emph{$\mathbb{Z}_2$ Topological Term, the Global Anomaly, and the Two-Dimensional Symplectic Symmetry Class of Anderson Localization,}
	Phys. Rev. Lett. {\bf 99}, 116601 (2007).
\bibitem{Ostrovsky2007} 
	P. M. Ostrovsky, I. V. Gornyi, and A. D. Mirlin, 
	\emph{Quantum Criticality and Minimal Conductivity in Graphene with Long-Range Disorder,}
	Phys. Rev. Lett. \textbf{98}, 256801 (2007).
\bibitem{Konig2014}
	E. J. K\"onig, P. M. Ostrovsky, I. V. Protopopov, I. V. Gornyi, I. S. Burmistrov, and A. D. Mirlin,
	\emph{Half-integer quantum Hall effect of disordered Dirac fermions at a topological insulator surface,}
	Phys. Rev. B {\bf 90}, 165435 (2014).
\bibitem{Ludwig1994} 
	A. W. W. Ludwig, M. P. A. Fisher, R. Shankar, G. Grinstein, 
	\emph{Integer quantum Hall transition: An alternative approach and exact results,}
	Phys. Rev. B, \textbf{50}, 7526 (1994).
\bibitem{Wang2014}
	C. Wang and T. Senthil, 
	\emph{Interacting fermionic topological insulators/superconductors in three dimensions,}
	Phys. Rev. B {\bf 89}, 195124 (2014). 
\bibitem{Foster2008}
	M. S. Foster and A. W. W. Ludwig,
	\emph{Metal-insulator transition from combined disorder and interaction effects in Hubbard-like electronic lattice models with random hopping,}
	Phys. Rev. B {\bf 77}, 165108 (2008).
\bibitem{Hosur2010}
	P. Hosur, S. Ryu, and Ashvin Vishwanath,	
	\emph{Chiral topological insulators, superconductors, and other competing orders in three dimensions,}
	Phys. Rev. B {\bf 81}, 045120 (2010).
\bibitem{TSCRev2}
	C.-K. Chiu, J. C. Y. Teo, A. P. Schnyder, and S. Ryu,
	\emph{Classification of topological quantum matter with symmetries,}
	Rev. Mod. Phys. {\bf 88}, 035005 (2016).
\bibitem{Nersesyan1994} 
	A. A. Nersesyan, A. M. Tsvelik, F. Wenger, 
	\emph{Disorder Effects in Two-Dimensional $d$-Wave Superconductors,}
	Phys. Rev. Lett. \textbf{72}, 2628, (1994).
\bibitem{Mudry1996}
	C. Mudry, C. Chamon, and X.-G. Wen, 
	\emph{Two-dimensional conformal field theory for disordered systems at criticality,}
	Nucl. Phys. B {\bf 466}, 383 (1996). 
\bibitem{Caux1996}
	J.-S. Caux, I. I. Kogan, and A. M. Tsvelik, 
	\emph{Logarithmic operators and hidden continuous symmetry in critical disordered models,}
	Nucl. Phys. B {\bf 466}, 444 (1996). 
\bibitem{Bhaseen2001}
	M. J. Bhaseen, J.-S. Caux, I. I. Kogan, and A. M. Tsvelik,
	\emph{Disordered Dirac fermions: the marriage of three different approaches,}
	Nucl. Phys. B {\bf 618}, 465 (2001).
\bibitem{AltlandSimonsZirnbauer2002}
	A. Altland, B. D. Simons, and M. R. Zirnbauer,
	\emph{Theories of Low-Energy Quasi-Particle States in Disordered $d$-Wave Superconductors,}
	Phys. Rep. {\bf 359}, 283 (2002).
\bibitem{DellAnna2007}
	L. Dell'Anna, M. Fabrizio, and Claudio Castellani,
	\emph{The role of the impurity-potential range in disordered d-wave superconductors,}
	J. Stat. Mech. {\bf P02014} (2007).
\bibitem{Chamon1996}
	C. C. Chamon, C. Mudry, and X.-G. Wen, 
	\emph{Localization in Two Dimensions, Gaussian Field Theories, and Multifractality,}
	Phys. Rev. Lett. {\bf 77}, 4194 (1996).
\bibitem{Tsvelik1995}
	A. M. Tsvelik, 
	\emph{Exactly solvable model of fermions with disorder,}
	Phys. Rev. B {\bf 51}, 9449 (1995).
\bibitem{Ostrovsky2006}
	P. M. Ostrovsky, I. V. Gornyi, and A. D. Mirlin, 
	\emph{Electron transport in disordered graphene,}
	Phys. Rev. B {\bf 74}, 235443 (2006).
\bibitem{Xie2015}
	H.-Y. Xie, Y.-Z. Chou, and M. S. Foster,
	\emph{Surface transport coefficients for three-dimensional topological superconductors,}
	Phys. Rev. B {\bf 91}, 024203 (2015).
\bibitem{Huckestein1995}
	B. Huckestein,
	\emph{Scaling theory of the integer quantum Hall effect,}
	Rev. Mod. Phys. {\bf 67}, 357 (1995).
\bibitem{Huo1993}
	Y. Huo, R. E. Hetzel, and R. N. Bhatt,
	\emph{Universal conductance in the lowest Landau level,} 
	Phys. Rev. Lett. {\bf 70}, 481 (1993).
\bibitem{Cho1997}	
	S. Cho and M. P. A. Fisher, 
	\emph{Conductance fluctuations at the integer quantum Hall plateau transition,}
	Phys. Rev. B {\bf 55}, 1637 (1997).
\bibitem{Wang1998}	
	X. Wang, Q. Li, and C. M. Soukoulis, 
	 \emph{Scaling properties of conductance at integer quantum Hall plateau transitions,}
	Phys. Rev. B {\bf 58}, 3576 (1998).
\bibitem{Jovanovic1998}
	B. Jovanovi\'c and Z. Wang,
	\emph{Conductance Correlations near Integer Quantum Hall Transitions,}
	Phys.\ Rev.\ Lett.\ {\bf 81}, 2767 (1998).
\bibitem{Schweitzer2005}
	L. Schweitzer and P. Marko\v s
	\emph{Universal Conductance and Conductivity at Critical Points in Integer Quantum Hall Systems,}
	Phys. Rev. Lett. {\bf 95}, 256805 (2005).
\bibitem{Chou2014} 
	Y.-Z. Chou and M. S. Foster, 
	\emph{Chalker scaling, level repulsion, and conformal invariance in critically delocalized quantum matter:
	Disordered topological superconductors and artificial graphene,}
	Phys. Rev. B {\bf 89}, 165136 (2014).
\bibitem{Ghorashi2018}
	S. A. A. Ghorashi, Y. Liao, and M. S. Foster,
	\emph{Critical Percolation without Fine Tuning on the Surface of a Topological Superconductor,}
	Phys.\ Rev.\ Lett.\ {\bf 121}, 016802 (2018).
\bibitem{Ghorashi2019}
	S. A. A. Ghorashi and M. S. Foster,
	\emph{Criticality Across the Energy Spectrum from Random, Artificial Gravitational Lensing in Two-Dimensional Dirac Superconductors,}
	arXiv:1903.11086. 
\bibitem{SQHPT-1_Kagalovsky1999}
	V. Kagalovsky, B. Horovitz, Y. Avishai, and J. T. Chalker,
	\emph{Quantum Hall Plateau Transitions in Disordered Superconductors,}
	Phys. Rev. Lett. {\bf 82}, 3516 (1999).
\bibitem{SQHPT-2_Gruzberg1999}
	I. A. Gruzberg, A.W.W. Ludwig, and N. Read, 
	\emph{Exact Exponents for the Spin Quantum Hall Transition,} 
	Phys. Rev. Lett. {\bf 82}, 4524 (1999).
\bibitem{SQHPT-3_Senthil1999}
	T. Senthil, J. B. Marston, and M. P. A. Fisher, 
	\emph{Spin quantum Hall effect in unconventional superconductors,} 
	Phys. Rev. B {\bf 60}, 4245 (1999).
\bibitem{SQHPT-4_Cardy2000}
	J. Cardy, 
	\emph{Linking Numbers for Self-Avoiding Loops and Percolation: Application to the Spin Quantum Hall Transition,}
	Phys. Rev. Lett. {\bf 84}, 3507 (2000).
\bibitem{SQHPT-5_Beamond2002}
	E. J. Beamond, J. Cardy, and J. T. Chalker, 
	\emph{Quantum and classical localization, the spin quantum Hall effect, and generalizations,} 
	Phys. Rev. B {\bf 65}, 214301 (2002).
\bibitem{SQHPT-6_Evers2003}
	F. Evers, A. Mildenberger, and A. D. Mirlin, 
	\emph{Multifractality at the spin quantum Hall transition,}
	Phys. Rev. B \textbf{67}, 041303R (2003).
\bibitem{SQHPT-7_Mirlin2003}
	A. D. Mirlin, F. Evers, and A. Mildenberger, 
	\emph{Wavefunction statistics and multifractality at the spin quantum Hall transition,}
	J. Phys. A \textbf{36}, 3255 (2003).
\bibitem{TSCRev3}
	M. Sato and Y. Ando,
	\emph{Topological superconductors: a review,}
	Rep. Prog. Phys. {\bf 80}, 076501 (2017).
\bibitem{D-1_SenthilFisher2000}
	T. Senthil and M. P. A. Fisher,
	\emph{Quasiparticle localization in superconductors with spin-orbit scattering,}
	Phys. Rev. B {\bf 61}, 9690 (2000).
\bibitem{D-2_BocquetZirnbauer2000}
	M. Bocquet, D. Serban, and M. R. Zirnbauer,
	\emph{Disordered 2d quasiparticles in class D:
	Dirac fermions with random mass, and dirty superconductors,}
	Nucl. Phys. B {\bf 578}, 628 (2000).
\bibitem{D-3_ReadLudwig2000}
	N. Read and A. W. W. Ludwig,
	\emph{Absence of a metallic phase in random-bond Ising models in two dimensions: 
	Applications to disordered superconductors and paired quantum Hall states,}
	Phys. Rev. B {\bf 63}, 024404 (2000). 
\bibitem{D-4_Gruzberg2001}
	I. A. Gruzberg, N. Read, and A. W. W. Ludwig,
	\emph{Random-bond Ising model in two dimensions:
	The Nishimori line and supersymmetry,}
	Phys. Rev. B {\bf 63}, 104422 (2001).
\bibitem{D-5_Chalker2001}
	J. T. Chalker, N. Read, V. Kagalovsky, B. Horovitz, Y. Avishai, and A. W. W. Ludwig,
	\emph{Thermal metal in network models of a disordered two-dimension superconductor,} 
	Phys. Rev. B {\bf 65}, 012506 (2001). 
\bibitem{D-6_Mildenberger2007}
	A. Mildenberger, F. Evers, A. D. Mirlin, and J. T. Chalker, 
	\emph{Density of quasiparticle states for a two-dimensional disordered system:
	Metallic, insulating, and critical behavior in the class-D thermal quantum Hall effect,}
	Phys. Rev. B {\bf 75}, 245321 (2007). 
\bibitem{D-7_Laumann2012}
	C. R. Laumann, A. W. W. Ludwig, D. A. Huse, and S. Trebst,
	\emph{Disorder-induced Majorana metal in interacting non-Abelian anyon systems,}
	Phys. Rev. B {\bf 85}, 161301(R) (2012).
\bibitem{SigmaModelManifoldNote}
	The nonlinear sigma models that describe the zero-energy class CI, AIII, and DIII
	2D topological surface states are Wess-Zumino-Novikov-Witten theories \cite{SRFL2008,Foster2014} 
	with $G \otimes G$ symmetry; for fermionic replicas, $G \in \{\text{Sp}(4n),\text{U}(2n),\text{O}(2n)\}$
	(Table~\ref{table:10-fold-way}). 
	As explained in Sec.~\ref{sec:sigma}, a finite-frequency perturbation to the sigma model
	breaks this down to diagonal $G$ symmetry. 
	For each class CI, AIII, and DIII, the relevant frequency perturbation is expected
	to drive the system to another sigma model with lower target manifold symmetry $G/H$. 
	As can be seen from the last column of Table~\ref{table:10-fold-way}, 
	this criterion is \emph{insufficient} to distinguish the ``exotic stacking'' and conventional
	scenarios, since (e.g.)  
	CI $\rightarrow$ C 
	vs.\
	CI $\rightarrow$ AI
	differ only by $H$. 
	The class CI, AIII, DIII sigma models without the WZNW term describe purely two-dimensional
	electron or superconductor quasiparticle systems; one expects the conventional flow to the associated
	Wigner-Dyson classes (and concomitant Anderson localization or weak antilocalization) at finite energy \cite{Evers2008}.
	The topological surface states can however behave differently,
	because they are anomalous bound states that cannot be detached from the higher-dimensional bulk \cite{Ghorashi2018}.  
	The WZNW term can be deformed into the Pruisken theta term for classes C, A, D in the exotic stacking scenario,
	although this argument appears insufficient to determine the theta coefficient (Sec.~\ref{sec:sigma}).
\bibitem{Cardy2013}
	J. Cardy,
	\emph{Logarithmic conformal field theories as limits of ordinary CFTs and some physical applications,}
	J. Phys. A: Math. Theor. {\bf 46}, 494001 (2013). 
\bibitem{Zirnbauer2019}
	M. R. Zirnbauer,
	\emph{The integer quantum Hall plateau transition is a current algebra after all,}
	Nucl. Phys. {\bf B941}, 458 (2019).
\bibitem{Slevin2009}
	K. Slevin and T. Ohtsuki,
	\emph{Critical exponent for the quantum Hall transition,}
	Phys. Rev. B {\bf 80}, 041304(R) (2009).
\bibitem{Obuse2010}
	H. Obuse, A. R. Subramaniam, A. Furusaki, I. A. Gruzberg, and A. W. W. Ludwig,
	\emph{Conformal invariance, multifractality, and finite-size scaling at Anderson localization transitions in two dimensions,}
	Phys. Rev. B {\bf 82}, 035309 (2010).
\bibitem{Amado2011}
	M. Amado, A. V. Malyshev, A. Sedrakyan, and F. Dom\'inguez-Adame,
	\emph{Numerical Study of the Localization Length Critical Index in a Network Model of Plateau-Plateau Transitions in the Quantum Hall Effect,}
	Phys. Rev. Lett. {\bf 107}, 066402 (2011).
\bibitem{Dahlhaus2011}
	J. P. Dahlhaus, J. M. Edge, J. Tworzydlo, and C. W. J. Beenakker, 
	\emph{Quantum Hall effect in a one-dimensional dynamical system,}
	Phys. Rev. B {\bf 84}, 115133 (2011).
\bibitem{Fulga2011}
	I. C. Fulga, F. Hassler, A. R. Akhmerov, and C. W. J. Beenakker,
	\emph{Topological quantum number and critical exponent from conductance fluctuations at the quantum Hall plateau transition,}
	Phys. Rev. B {\bf 84}, 245447 (2011).
\bibitem{Obuse2012}
	H. Obuse, I. A. Gruzberg, and F. Evers,
	\emph{Finite-Size Effects and Irrelevant Corrections to Scaling Near the Integer Quantum Hall Transition,}
	Phys. Rev. Lett. {\bf 109}, 206804 (2012).
\bibitem{Slevin2012}
	K. Slevin and T. Ohtsuki, 
	\emph{Finite size scaling of the Chalker-Coddington model,}
	Int. J. Mod. Phys. Conf. Ser. {\bf 11}, 60 (2012).
\bibitem{Nuding2015}
	W. Nuding, A. Kl\"umper, and A. Sedrakyan, 
	\emph{Localization length index and subleading corrections in a Chalker-Coddington model: A numerical study,}
	Phys. Rev. B {\bf 91}, 115107 (2015).
\bibitem{Gruzberg2017}
	I. A. Gruzberg, A. Kl\"umper, W. Nuding, and A. Sedrakyan,
	\emph{Geometrically disordered network models, quenched quantum gravity, and critical behavior at quantum Hall plateau transitions,}
	Phys. Rev. B {\bf 95}, 125414 (2017).
\bibitem{Foster2014}
	M. S. Foster, H.-Y. Xie, and Y.-Z. Chou, 
	\emph{Topological protection, disorder, and interactions: Survival at the surface of 3D topological superconductors,}
	Phys. Rev. B {\bf 89}, 155140 (2014).
\bibitem{GadeWegner1991}
	R. Gade and F. Wegner, 
	\emph{The $n = 0$ replica limit of U($n$) and U($n$)/SO($n$) models,}
	Nucl. Phys. {\bf B360}, 213 (1991).
\bibitem{Gade1993}
	R. Gade,
	\emph{Anderson localization for sublattice models,}
	Nucl. Phys. {\bf B398}, 499 (1993).
\bibitem{Guruswamy2000}
	S. Guruswamy, A. LeClair, and A. W. W. Ludwig,
	\emph{$gl(N|N)$ Super-current algebras for disordered Dirac fermions in two dimensions,}
	Nucl. Phys. {\bf B583}, 475 (2000).
\bibitem{AleinerEfetov2006}
	I. L. Aleiner and K. B. Efetov, 
	\emph{Effect of Disorder on Transport in Graphene,}
	Phys. Rev. Lett. {\bf 97}, 236801 (2006).
\bibitem{FosterAleiner2008}
	M. S. Foster and I. L. Aleiner,
	\emph{Graphene via large N: A renormalization group study,}
	Phys. Rev. B {\bf 77}, 195413 (2008).
\bibitem{GrapheneReview2011}
	V. N. Kotov, B. Uchoa, V. M. Pereira, F. Guinea, and A. H. Castro Neto,
	\emph{Electron-Electron Interactions in Graphene: Current Status and Perspectives,}
	Rev. Mod. Phys. {\bf 84}, 1067 (2012).
\bibitem{Peres2010}
	N. M. R. Peres,
	\emph{Colloquium: The transport properties of graphene: An introduction,}
	Rev. Mod. Phys. {\bf 82}, 2673 (2010).
\bibitem{Ho1998}
	T.-L. Ho,
	\emph{Spinor Bose Condensates in Optical Traps,}
	Phys. Rev. Lett. {\bf 81}, 742 (1998).
\bibitem{Ohmi1998}
	T. Ohmi and K. Machida, 
	\emph{Bose-Einstein Condensation with Internal Degrees of Freedom in Alkali Atom Gases,}
	J. Phys. Soc. Jpn., {\bf 67}, 1822 (1998). 
\bibitem{Koenig2012}
	E. J. K\"onig, P. M. Ostrovsky, I. V. Protopopov, and A. D. Mirlin,
	\emph{Metal-insulator transition in two-dimensional random fermion systems of chiral symmetry classes,}
	Phys. Rev. B {\bf 85}, 195130 (2012).
\bibitem{BernardLeClair2002}
	D. Bernard and A. LeClair, 
	\emph{A classification of 2D random Dirac fermions,}
	J. Phys. A {\bf 35}, 2555 (2002).
\bibitem{Castillo1997}
	H. E. Castillo, C. C. Chamon, E. Fradkin, P. M. Goldbart, and C. Mudry, 
	\emph{Exact calculation of multifractal exponents of the critical wave function of Dirac fermions in a random magnetic field,}	
	Phys. Rev. B {\bf 56}, 10668 (1997).
\bibitem{Carpentier2001}
	D. Carpentier and P. Le Doussal,
	\emph{Glass transition of a particle in a random potential, front selection in nonlinear renormalization group, and entropic phenomena in Liouville and sinh-Gordon models,}
	Phys. Rev. E {\bf 63}, 026110 (2001).
\bibitem{Motrunich2002}
	O. Motrunich, K. Damle, and David A. Huse,
	\emph{Particle-hole symmetric localization in two dimensions,}
	Phys. Rev. B {\bf 65}, 064206 (2002). 
\bibitem{Mudry2003}
	C. Mudry, S. Ryu, and A. Furusaki, 
	\emph{Density of states for the $\pi$-flux state with bipartite real random hopping only: A weak disorder approach,}
	Phys. Rev. B {\bf 67}, 064202 (2003).
\bibitem{DellAnnaFreezing2006}
	L. Dell'Anna, 
	\emph{Anomalous dimensions of operators without derivatives in the non-linear sigma model for disordered bipartite lattices,} 
	Nucl. Phys. {\bf B750}, 213 (2006).
\bibitem{Evers2014}
	V. H\"afner, J. Schindler, N. Weik, T. Mayer, S. Balakrishnan, R. Narayanan, S. Bera, and F. Evers,
	\emph{Density of States in Graphene with Vacancies: Midgap Power Law and Frozen Multifractality,}
	Phys. Rev. Lett. {\bf 113}, 186802 (2014).
\bibitem{Tworzydlo2006}
	J. Tworzydlo, B. Trauzettel, M. Titov, A. Rycerz, and C. W. J. Beenakker, 
	\emph{Sub-Poissonian Shot Noise in Graphene,}
	Phys. Rev. Lett. {\bf 96}, 246802 (2006).
\bibitem{DellAnna2006}
	L. Dell'Anna, 
	\emph{Disordered d-wave superconductors with interactions,}
	Nucl. Phys. {\bf B758}, 255 (2006).
\bibitem{Schuessler2009}
	A. Schuessler, P. M. Ostrovsky, I. V. Gornyi, and A. D. Mirlin,
	\emph{Analytic theory of ballistic transport in disordered graphene,}
	Phys. Rev. B {\bf 79}, 075405 (2009).
\bibitem{FeastAlgo} 
	E. Polizzi,
	\emph{Density-Matrix-Based Algorithms for Solving Eigenvalue Problems,}
	Phys. Rev. B. {\bf 79}, 115112 (2009).
\bibitem{Evers2001}
	F. Evers, A. Mildenberger, and A. D. Mirlin
	\emph{Multifractality of wave functions at the quantum Hall transition revisited,}
	Phys. Rev. B {\bf 64} 241303(R) (2001).
\bibitem{Pruisken1987}
	A. M. M. Pruisken, 
	\emph{Field Theory, Scaling and the Localization Problem,} 
	in 
	\emph{The Quantum Hall Effect}, 
	edited by R. E. Prange and S. M. Girvin 
	(Springer-Verlag, New York, 1987). 
\bibitem{FradkinBook}
	E. Fradkin,
	\emph{Field Theories of Condensed Matter Systems}
	(Westview Press, Boulder, Colorado, 1991).
\end{thebibliography}
\end{document}